\documentclass[11pt]{article}
\usepackage{geometry}                
\usepackage{amsmath, amsthm, amssymb}
\usepackage{epsfig}

\usepackage{graphicx}
\graphicspath {{figs/}}

\usepackage{epstopdf}
\usepackage{enumerate}
\usepackage{subfig}
\usepackage[english]{babel}
\usepackage{longtable}
\usepackage{xcolor}
\usepackage[margin=1cm]{caption}

\newtheorem{theorem}{Theorem}[section]

\newtheorem{proposition}[theorem]{Proposition}

\theoremstyle{definition}
\newtheorem{definition}[theorem]{Definition}

\theoremstyle{remark}

\newcommand{\1}{\mathbf{1}}
\newcommand{\0}{\mathbf{0}}
\newcommand{\bZ}{\mathbf{Z}}
\newcommand{\Tr}{\mathrm{Tr}}
\newcommand{\Pf}{\mathrm{Pf}}

\newcommand{\erfc}{\mathrm{erfc}}
\newcommand{\qdet}{\mathrm{qdet}}

\newcommand{\diag}{\mathrm{diag}}

\newcommand{\bX}{\mathbf{X}}
\newcommand{\bR}{\mathbf{R}}
\newcommand{\bA}{\mathbf{A}}
\newcommand{\bB}{\mathbf{B}}
\newcommand{\bC}{\mathbf{C}}
\newcommand{\bM}{\mathbf{M}}

\newcommand{\bT}{\mathbf{T}}

\newcommand{\bQ}{\mathbf{Q}}

\newcommand{\bU}{\mathbf{U}}

\newcommand{\bD}{\mathbf{D}}

\newcommand{\bG}{\mathbf{G}}

\newcommand{\bY}{\mathbf{Y}}
\newcommand{\bK}{\mathbf{K}}

\newcommand{\bx}{\mathbf{x}}
\newcommand{\bv}{\mathbf{v}}
\newcommand{\bz}{\mathbf{z}}
\newcommand{\cc}[1]{\overline{#1}}

\definecolor{darkred}{rgb}{0.6,0,0}

\DeclareMathOperator{\e}{e}

\begin{document}

\title{An induced real quaternion spherical ensemble of random matrices}
\author{Anthony Mays \thanks{School of Mathematics and Statistics, University of Melbourne, Australia.\newline \emph{Email}: anthony.mays@unimelb.edu.au}
\and Anita Ponsaing \thanks{ARC Centre of Excellence for Mathematical and Statistical Frontiers (ACEMS), School of Mathematics and Statistics, University of Melbourne, Australia.\newline \emph{Email}: aponsaing@unimelb.edu.au}
}
\date{\today}

\maketitle



\begin{abstract}
We study the induced spherical ensemble of non-Hermitian matrices with real quaternion entries (considering each quaternion as a $2\times 2$ complex matrix). We define the ensemble by the matrix probability distribution function that is proportional to
\begin{align*}
\frac{\det (\bG \bG^{\dagger})^{2 L}} {\det(\1_N+ \bG \bG^{\dagger})^{2 (n+N+L)}}.
\end{align*}
These matrices can also be constructed via a procedure called `inducing', using a product of a Wishart matrix (with parameters $n,N$) and a rectangular Ginibre matrix of size $(N+L)\times N$. The inducing procedure imposes a repulsion of eigenvalues from $0$ and $\infty$ in the complex plane, with the effect that in the limit of large matrix dimension, they lie in an annulus whose inner and outer radii depend on the relative size of $L$, $n$ and $N$.

By using functional differentiation of a generalized partition function, we make use of skew-orthogonal polynomials to find expressions for the eigenvalue $m$-point correlation functions, and in particular the eigenvalue density (given by $m=1$).

We find the scaled limits of the density in the bulk (away from the real line) as well as near the inner and outer annular radii, in the four regimes corresponding to large or small values of $n$ and $L$. After a stereographic projection the density is uniform on a spherical annulus, except for a depletion of eigenvalues on a great circle corresponding to the real axis (as expected for a real quaternion ensemble). We also form a conjecture for the behaviour of the density near the real line based on analogous results in the $\beta=1$ and $\beta=2$ ensembles; we support our conjecture with data from Monte Carlo simulations of a large number of matrices drawn from the $\beta=4$ induced spherical ensemble.
\end{abstract}

\section{Introduction and main results}

Non-Hermitian random matrices largely began with the pioneering work of Ginibre in 1965 \cite{Gini1965}, which discussed three ensembles of matrices having independent real, complex and real quaternion\footnote{When we say `real quaternion' we mean a quaternion in the sense of a number $q= q_0 +i q_1 +j q_2 + k q_3$, where $q_m \in \mathbb{R}$ and $q$ obeys the quaternionic multiplication and addition rules. These real quaternions can be represented as $2\times 2$ complex matrices and it is in this representation that we calculate the (complex) eigenvalues of real quaternionic matrices. We provide an overview of quaternionic definitions and properties in Appendix \ref{app:quatPf}.} random entries respectively, in keeping with Dyson's three-fold way \cite{Dyso1962}. As with Hermitian ensembles, these non-Hermitian ensembles correspond to the indices $\beta=1, 2, 4$ respectively, which represent the number of independent real components in each matrix entry.

More recently, various other non-Hermitian ensembles have attracted interest (see \cite{Akem2005, ForrNaga2007, KhorSommZycz2010, KhorSomm2011, AkemPhil2014} for a small selection). One particular categorization relevant to the present work is the `geometrical triumvirate' of ensembles described in \cite{Kris2006, HougKrisPereVira2009, Mays2011, Fisc2013}, which identifies random matrix ensembles with the three classical surfaces of constant curvature: the plane, the sphere and the pseudo- or anti-sphere. We leave the interested reader to investigate for themselves all the details contained in those works; here we highlight only the spherical ensemble, which is given by the matrix `ratio'
\begin{align}
\label{e:YAB} \bY^{\vphantom{-1}} = \bA^{-1} \bB^{\vphantom{-1}},
\end{align}
where $\bA$ and $\bB$ are independent $N\times N$ Gaussian matrices (\textit{i.e.}, they are drawn from the Ginibre ensembles, which correspond to the plane), and $\bA$ is non-singular. By analogy with Cauchy random variables (which can be described as the ratio of two Gaussian random variables) these matrices have been called \textit{Cauchy matrices} \cite{EdelKostShub1994}, and have the matrix Cauchy distribution function \cite{Fein2004, HougKrisPereVira2009, ForrMays2011, Mays2013}
\begin{align}
\label{e:sphmpdf} \mathcal{P}^{(\beta)} (\bY) = \pi^{-\beta N^2/2} \prod_{j=0}^{N-1} \frac{\Gamma \left( \frac{N+1+j} {2}\beta\right)} {\Gamma \left( \frac{j+1}{2}\beta \right)} \det (1+\bY \bY^{\dagger} )^{-\beta N},
\end{align}
where the `dagger' should be interpreted as `transpose', `Hermitian conjugate' or `quaternion dual' for $\beta =1$ (real matrices), $\beta =2$ (complex matrices) and $\beta =4$ (real quaternion matrices) respectively. Note that for $\beta=4$, the determinant is to be understood as a quaternion determinant (see Appendix \ref{app:quatPf}). We note also that the eigenvalues of the matrix $\bY$ defined in \eqref{e:YAB} are equal to the \textit{generalized eigenvalues} of the pair $(\bA, \bB)$, given as the solutions $\lambda_j$ to the equation
\begin{align}
\nonumber \det(\bB- \lambda_j \bA)=0,
\end{align}
which is the viewpoint of \cite{EdelKostShub1994}.

As in the case of the Ginibre ensembles, the eigenvalue density has distinctive symmetries depending on the value of $\beta$:
\begin{itemize}
\item{$\beta=2$ ($\bA$ and $\bB$ complex): rotational symmetry in the complex plane;}
\item{$\beta=1$ ($\bA$ and $\bB$ real): positive density of eigenvalues along the real axis, with reflective symmetry across the real axis;}
\item{$\beta=4$ ($\bA$ and $\bB$ real quaternion): depletion of eigenvalues near the real axis, with reflective symmetry across the real axis.}
\end{itemize}
(The reflective symmetry is a property of all finite-size real and real quaternion matrices, where the non-real eigenvalues come in complex-conjugate pairs.) The reason for the name `spherical ensemble' is that the eigenvalues of $\bY$ have uniform distribution (under stereographic projection) on the unit sphere in the limit of large matrix dimension, which is a consequence of the \textit{spherical law} \cite{Roge2010, ForrMays2011, Bord2011}, a result analogous to the more famous \textit{circular law} for Ginibre matrices (see for example \cite{Girk1984, Bai1997, GoetTikh2010, TaoVuKris2010}). For details concerning the eigenvalue statistics of these ensembles the interested reader may refer to \cite{HougKrisPereVira2009, Mays2011}, in addition to the works listed above. One may also seek a physical interpretation of these processes in terms of minimizing some energy function on a sphere, in which case we refer the reader to \cite{LeCaHo1990, ArmeBeltShub2011, BoroSerf2013}.

Each class in the geometrical triumvirate can be generalized by the introduction of parameters whose effect is to restrict the eigenvalue density to an annulus in the complex plane through a procedure called `inducing' (we provide a brief overview of this procedure in Appendix \ref{s:genrand}, but comprehensive descriptions are given in \cite{FBKSZ2012, FiscForr2011, Fisc2013}). However, we take as our definition of the \textit{induced spherical ensemble} those $N\times N$ matrices $\bG$ that are defined by the matrix probability density function (pdf)
\begin{align}
\label{d:sphmats} \mathcal{P}^{(\beta)} (\bG) := K_N^{(\beta)} \frac{\det (\bG \bG^{\dagger})^{\beta L /2}} {\det(\1_N+ \bG \bG^{\dagger})^{\beta (n+N+L)/2}},
\end{align}
where $n \geq N$ and $L\geq 0$; as mentioned above the parameter $\beta$ corresponds to matrices with real ($\beta=1$), complex ($\beta=2$) or real quaternion ($\beta=4$) entries. The normalization constant $K_N^{(\beta)}$ is given by 
\begin{align}
\label{e:KN} K_N^{(\beta)} = \pi^{-\beta N^2/2} \prod_{j=1}^N \frac{\Gamma(\beta j/2) \Gamma(\beta (n+L+j)/2)} {\Gamma(\beta (L+j)/2) \Gamma(\beta (n- N+j)/2)},\qquad \beta=1,2,4.
\end{align}
The ensemble corresponding to $\beta= 2$ was the subject of \cite{FiscForr2011}, while that corresponding to $\beta= 1$ was discussed in \cite{Fisc2013}; the aim of the work in this paper is to study the eigenvalue statistics of the analogous $\beta=4$ real quaternion ensemble. First note that with $L\to 0$, $n\to N$, (\ref{d:sphmats}) reduces to (\ref{e:sphmpdf}) and we are back in the regime of the spherical law, which was mentioned above. The result of the generalization (\ref{d:sphmats}) is to keep the eigenvalues away from the origin and $\infty$, effectively squeezing the support into an annulus. This annulus projects (stereographically) to a belt of eigenvalues centered on the great circle corresponding to the circle $|z|=1$. As an aid to visualization in Figures \ref{f:quatBB}--\ref{f:quatSS} of Section \ref{s:slims} we present simulated eigenvalue plots for $\beta= 4$ and their stereographic projections onto the unit sphere. In brief, as in the $\beta=1$ and $\beta=2$ cases we find four regimes that correspond to large and small values (compared to $N$) of $L$ and $n$. Although we take the pdf \eqref{d:sphmats} to be our definition of the induced spherical matrices, as alluded to above, it is possible to explicitly construct them from products of Wishart and Ginibre matrices. While the real quaternion construction is a natural modification of the discussions in the above references, there are some subtleties related to numerical Monte Carlo simulations of the real quaternion induced spherical ensemble and so we make some comments on this point in Appendix \ref{s:genrand}.

We note that these matrices are similar to the class of matrices that relate to the Feinberg--Zee \textit{single ring theorem}, which was discussed in \cite{FeinZee1997} and rigorously proved in \cite{GuioKrisZeit2009}. The theorem states that for $N\times N$ complex matrices $\phi$ from a distribution
\begin{align}
\label{e:FZ} \mathcal{P}(\phi) = \frac{1} {Z_N} e^{-N \;\Tr \: V (\phi \phi^{\dagger})},
\end{align}
where $V(\phi \phi^{\dagger})$ is a polynomial with positive leading coefficient, the support of the eigenvalue density tends toward an annulus around the origin, and the density is rotationally symmetric. From the figures in Section \ref{s:slims} we see that the eigenvalue densities certainly have these properties, yet (\ref{d:sphmats}) is a special case of (\ref{e:FZ}) only formally (in the sense that we require $V$ to be a general analytic function). More work is needed to make this connection precise.

The explicit goal of the present work is to calculate the eigenvalue correlation functions and various scaled limits of the eigenvalue density for the real quaternion matrices drawn from the distribution (\ref{d:sphmats}), which therefore generalizes the results in \cite{Mays2013}. As mentioned above, the complex analogue of this work was presented in \cite{FiscForr2011, Fisc2013} while the real case can be found in \cite{Fisc2013}. Since quaternions, quaternion determinants and Pfaffians play a crucial role in this work we provide a review in Appendix \ref{app:quatPf}.

To obtain our results we will make use of a \textit{generalized partition function}, which, for a general joint probability density function (jpdf) $\mathcal{Q} (\bz_N)$ in $N$ variables $z_1, \dots, z_N$, is defined by the average
\begin{align}
\label{e:gpf} Z_N [\bv]:=\left \langle \prod_{j=1}^N v_j (z_j) \right\rangle_{\mathcal{Q}} = \int_{\Omega} dz_{1} \; v_1 (z_1) \cdots \int_{\Omega} dz_{N} \; v_N (z_N) \: \mathcal{Q} (\bz_N),
\end{align}
where $\bv= \{ v_1, \dots, v_N \}$ are some well-behaved functions in the variables $z_j \in \Omega$. In \cite{Sinc2007} it was shown that (\ref{e:gpf}) can be written in a convenient Pfaffian form for various eigenvalue jpdfs, of which the one considered in this work is an example. This allows us to follow \cite{ForrNaga2007, BoroSinc2009} and use (\ref{e:gpf}) to calculate the eigenvalue correlation functions. For a general jpdf $\mathcal{Q} (\bz_N)$ the \textit{$m$-point correlation function} is defined by
\begin{align*}
\rho_{(m)}(r_1, \dots, r_m)&:=\frac{N(N-1) \cdots (N-m+1)} {Z_N [1]} \int_{\Omega} dz_{m+1} \cdots \int_{\Omega} dz_N \\
&\qquad\times  \mathcal{Q} (r_1, \dots, r_m, z_{m+1} ,\dots ,z_N),
\end{align*}
in terms of which the eigenvalue density is given by $\rho_{(1)} (x)$, with the normalization
\begin{align}
\label{e:edens} \int_{\Omega} \rho_{(1)} (x) dx =N.
\end{align}
Equivalently we can obtain the correlation functions via functional differentiation of the generalized partition function
\begin{align}
\label{e:correlnd} \rho_{(m)} (r_1,  \dots , r_m)= &\left. \frac{1}{Z_N [\bv]} \frac{\delta^{m}} {\delta v_1 (z_1) \cdots \delta v_m (z_m)} Z_N [\bv] \right|_{\bv =\1}.
\end{align}
We will find that for the jpdf we consider in this work, $Z_N [\bv]$ can be written as a Fredholm Pfaffian (or quaternion determinant), which via (\ref{e:correlnd}) yields the correlation functions immediately (see Section \ref{s:gpf}).

Our method here falls into the category of (skew-)orthogonal polynomial methods and, as such, we will have need of the polynomials corresponding to the generalized partition functions (\ref{e:gpf}). It is not yet known how to complete a calculation analogous to that in \cite{Fisc2013} for $\beta =1$, where the skew-orthogonal polynomials are deduced directly from an average over characteristic polynomials, however, a result from \cite{Forr2013} furnishes us with the necessary expressions. Armed with these polynomials we establish the eigenvalue correlation functions in Propositions \ref{p:correlns4}. From these correlation functions we find (with $\mathrm{Im}(z) >0$) that the eigenvalue density (normalized according to (\ref{e:edens})) is
\begin{align}
\nonumber \rho_{(1)} (z) &=\frac{i\;\Gamma (2n+2L+2)\; \mathrm{Im} (z) \; |z|^{4L}} {2^{2(L+n) -1}(1+|z|^2)^{2n+2L+2}}\\
\label{e:dens4}&\times \sum_{k=0}^{N-1} \sum_{j=0}^k \frac{|z|^{4j} \left(\cc{z}^{2k-2j+1} -  z^{2k-2j+1} \right)} {\Gamma (L+j+1) \Gamma (L+k+3/2) \Gamma (n-j+1/2) \Gamma (n-k)}.
\end{align}

Having established the correlation functions for finite matrix sizes $N$, we then analyze various scaled limits of the eigenvalue density in Section \ref{s:slims}. As discussed above, it is known from the spherical law that for spherical matrices (\ref{e:sphmpdf}) the eigenvalue density is uniform (under stereographic projection) on the unit sphere. The figures in Section \ref{s:slims} suggest the eigenvalue support is restricted to an annulus in the complex plane for large matrix size, the inner and outer radii of which depend on the relative sizes of $L$, $n$ and $N$. Indeed, as in \cite{Fisc2013}, we can identify four regimes of interest as $N \to \infty$: (i) $L= O (N)$, $n- N= O (N)$; (ii) $L= O (N)$, $n- N= O (1)$; (iii) $L= O (1)$, $n- N= O (N)$; and (iv) $L= O (1)$, $n- N= O (1)$. Broadly speaking, for large $L$ the eigenvalues are repulsed from the origin (which corresponds to the south pole), and for large $n- N$ the eigenvalues are repulsed from infinity (the north pole). While we find that we can calculate the limiting bulk and annular edge densities in these four regimes, we are not yet able to derive the density near the real line. We present a conjecture for this in Section \ref{s:gencorrelns4} along with some simulated data to support the claim. Further, we discuss a differential equation, which, if it was to be solved, should also yield the asymptotics for the full eigenvalue correlation function in this and similar ensembles --- however, based upon the structure of the equation (and similar difficulties in related studies, eg \cite{Ipse2015}) this appears a remote possibility.

\subsection{Some notational conventions}

To avoid confusion we state here some of the notation commonly used in this paper. We will usually use upper-case bold letters (e.g. $\bA, \bB$) to denote matrices, often with an accompanying subscript to denote the matrix dimension. We use the symbol $\dagger$ to refer to `transpose', `Hermitian conjugate' or `quaternion dual' for real, complex and real quaternion matrices, respectively; occasionally, in order to be clear on the matrix type, we will use the superscripts $T, \dagger$ and $D$ to denote them explicitly. A detailed description of the properties of the relevant quaternion properties is contained in Appendix \ref{app:quatPf}.

Lower case bold letters are lists (they need not be ordered), \textit{e.g.} $\bx_M = \{ x_1,  \dots , x_M \}$, where the subscript denotes the length. In particular, the bold $\boldsymbol{\lambda}_M$ will always denote the list of eigenvalues of a system of size $M$. Generally these eigenvalues will either be real or live in the upper half complex plane, that is $\mathbb{C}_{+}:=\{z\ |\ \mathrm{Im}(z)>0 \}$.

We will denote the wedge product of complex and real quaternion quantities respectively by $(dz)=dx\wedge dy$ for $z=x+ iy$ and $(dq)= dq_0 \wedge dq_1 \wedge dq_2 \wedge dq_3$ for $q=q_0+iq_1+jq_2+kq_3$. The wedge product of the differentials of the independent real entries of an object (a matrix or list) are then given by
\begin{align*}
(d\bA):= \bigwedge_{j,k} (da_{j,k}), \qquad (d\bx_M):= \bigwedge_j (dx_j),
\end{align*}
where the indices run over all values corresponding to independent elements.

We make use of the (half-max) Heaviside step function
\begin{align*}
\Theta(x):= \left\{ \begin{array}{cl}
0,& x<0,\\
1/2,& x =0,\\
1,& x> 0.
\end{array} \right.
\end{align*}

\section{Normalization of the matrix pdf}

The normalization in the complex case $K_N^{(2)}$ was presented in \cite{FiscForr2011} and the real case $K_N^{(1)}$ in \cite{Fisc2013}; by performing a similar procedure the normalization $K_N^{(4)}$ can also be calculated explicitly.

\begin{proposition}\label{p:KN}
With integers $L=M -N \geq 0$, $n\geq N> 0$ the matrix probability density function (\ref{d:sphmats}) for the real quaternion induced spherical ensemble has normalization $K^{(4)}_N$ given by (\ref{e:KN}).
\end{proposition}

\proof We search for $K^{(4)}_N$ such that
\begin{align}
\label{e:matpdfK4a} K^{(4)}_N \int \frac{\qdet( \hat{\bG}^D \hat\bG)^{2(M-N)}}{\qdet(\1_N+ \hat\bG^{D} \hat\bG)^{2(n+M)}} (d\hat\bG) = 1,
\end{align}
using the $1\times 1$ representation of the quaternion, according to the notation in Appendix \ref{app:quatPf} (where the superscript $D$ is the quaternion dual operation). Let $\hat\bC:= \hat\bG^D \hat\bG$, for which we have the Jacobian \cite{Olki2002}
\begin{align}
\nonumber (d\hat\bG) = \tilde{c}\; \qdet\: \hat\bC \; (d\hat\bC),
\end{align}
where $\tilde{c}$ is independent of $\hat\bG$, and (\ref{e:matpdfK4a}) becomes
\begin{align*}
 \frac1{K^{(4)}_N} &= \tilde{c} \int_{\hat\bC >0} \frac{\qdet( \hat\bC)^{2(M-N) +1}}{\qdet(\1_N+ \hat\bC)^{2(n+M)}} (d \hat\bC)\\
&= \tilde{c} \int (\bQ^{\dagger} d\bQ) \int_0^{\infty} d\lambda_1 \dots \int_0^{\infty} d\lambda_N \prod_{j=1}^N \frac{\lambda_j^{2(M-N)+1}} {(1+\lambda_j)^{2(n +M)}} \prod_{j<k} |\lambda_k- \lambda_j|^4,
\end{align*}
where $\bC=\bQ^{\dagger}\:\diag(\lambda_1,\lambda_1, \lambda_2, \lambda_2, \dots, \lambda_N, \lambda_N)\:\bQ$ is a unitary eigendecomposition of the $2\times 2$ block representation of $\hat\bC$, and similarly, $\bQ$ is such that $\hat\bQ\in Sp(N)/(U(1))^N$. For the second equality we have made use of the well-known Jacobian for changing variables from the matrix entries $\bC_{j, k}$ to the matrix eigenvalues $\lambda_j$ (see for example \cite[Chapter 1.3]{Forr2010}). Now replace $\lambda_j= t_j/(1-t_j)$, giving $d\lambda_j= dt_j/(1-t_j)^2$ and $1+\lambda_j= (1-t_j)^{-1}$, which leads to the Selberg integral \cite{Selb1944}
\begin{align}
\nonumber \frac1{K^{(4)}_N} &= \tilde{c} \int (\bQ^{\dagger} d\bQ) \int_0^{1} dt_1 \dots \int_0^{1} dt_N \prod_{j=1}^N t_j^{2L+1} (1-t_j)^{2n-2N+1} \prod_{j<k} |t_k- t_j|^4\\
\label{e:Kinv} &= \tilde{c} \prod_{j=1}^{N} \frac{\Gamma (2(L+j)) \Gamma (2(n-N+ j)) \Gamma (2j+1)} {\Gamma (2(L+n+j)) \Gamma(3)} \int (\bQ^{\dagger} d\bQ).
\end{align}
An evaluation of the integral over $(\bQ^\dagger d\bQ)$ can be found in \cite{Nach1965}, however it won't be necessary for our purposes. Using
\begin{align}
\nonumber \int e^{-(\Tr \bG \bG^{\dagger})/4} (d\bG) = \tilde{c} \int_{\bC >0} e^{-(\Tr \bC) /4} (\det \bC)^{1/2} (d\bC),
\end{align}
we can calculate $\tilde{c}$ as in \cite{Mays2013} and find
\begin{align}
\nonumber \tilde{c} &= \pi^{2N^2} \prod_{j=1}^N \frac{\Gamma (3)} {\Gamma(2j) \Gamma (2j+1)} \left( \int (\bQ^{\dagger} d\bQ) \right)^{-1}.
\end{align}
Substituting this into \eqref{e:Kinv} we have
\begin{align*}
 K^{(4)}_N= \pi^{-2N^2} \prod_{j=1}^N \frac{\Gamma (2j) \Gamma (2(L+n+j))} {\Gamma(2(L+j)) \Gamma (2(n-N+j))}.
\end{align*}

\hfill $\Box$

\section{Eigenvalue jpdf}
\label{s:ejpdf}

Here we change variables in the matrix pdf (\ref{d:sphmats}) to the eigenvalues for the real quaternion ensemble following the methods of \cite{Mays2013} (which deals with the specified ensemble $L\mapsto 0$, $n\mapsto N$). The idea is to use a Schur decomposition
\begin{align}
\nonumber \bG= \bU \bR \bU^{-1},
\end{align}
where $\bU$ is a symplectic matrix (\textit{i.e.}, a unitary real quaternion matrix) and $\bR$ is a (block) upper triangular matrix, whose diagonal blocks correspond to the eigenvalues $\boldsymbol{\lambda}$ of $\bG$. We have the relation
\[ \int \mathcal P(\bG)(d\bG) = \mathcal Q(\boldsymbol{\lambda})(d\boldsymbol{\lambda}) \]
between the matrix pdf $\mathcal P$ and the eigenvalue jpdf $\mathcal Q$, where the integral is understood to be over the variables relating to the eigenvectors. Performing the integral involves iteratively integrating column-by-column over the blocks in the strict upper triangle of $\bR$ (a technique introduced to this topic in \cite{HougKrisPereVira2009}) as well as an integral over $(\bU^\dag d\bU)$ \cite{Nach1965}. Except for the factors of $\lambda^{\beta L /2}$ coming from the numerator of (\ref{d:sphmats}) the procedure here is identical and so we will not include it in full; the interested reader is referred to \cite{Mays2013, Fisc2013}.

\begin{proposition}
With $z_j\in \mathbb{C}_+$, the eigenvalue jpdf for the real quaternion induced spherical ensemble is
\begin{align}
\label{e:ejpdfQ} \mathcal{Q} (\mathbf{z}_N, \mathbf{\cc{z}}_N) (d \mathbf{z}_N) &= \frac{C_{N}} {\Gamma (N+1)} \Delta (\mathbf{z}_N, \cc{\mathbf{z}}_N) \prod_{j=1}^N h (z_j) h (\cc{z}_j)\; dx_j dy_j,
\end{align}
where
\begin{align}
\nonumber h (z) &:= \frac{|z|^{2L} (z- \cc{z})^{1/2}} {(1+|z|^2)^{n+L+1}},\\
\nonumber C_{N} &:= \frac{(-1)^{N(N-1)/2}} {\pi^{N}} i^N \prod_{j=1}^N \frac{\Gamma (2n+2L+2)} {\Gamma (2L+2j) \Gamma (2n-2N+ 2j)}.
\end{align}
\end{proposition}
In the definition of $C_N$ above we have kept the factor of $i^N$ separate from the powers of $-1$ for clarity; this factor comes from splitting the factors $\prod_{j=1}^N (z- \cc{z})$ into $\prod_{j=1}^N h (z) h (\cc{z})$.

\section{Eigenvalue correlation functions}
\label{s:gpf}

As mentioned in the introduction, to find the eigenvalue correlation functions we will first find the generalized partition function (\ref{e:gpf}) and then use the functional differentiation formula (\ref{e:correlnd}) to obtain the correlation functions. We know from \cite{DeBr1955, Meht2004, Sinc2007} that pdfs of the form (\ref{e:ejpdfQ}) can be transformed to a more convenient Pfaffian or quaternion determinant form using the method of integration over alternate variables via the Vandermonde identity. We state only the results here; the interested reader is referred to \cite{Forr2010, Mays2011, Fisc2013} (in addition to those references mentioned above) for explicit details. (For the real quaternion ensemble we take $v_1= \dots = v_N = v$ in (\ref{e:gpf}).)
\begin{proposition}
\label{p:gpfQ}
The generalized partition function for the real quaternion induced spherical ensemble with eigenvalue jpdf (\ref{e:ejpdfQ}) is
\begin{align}
\label{e:ZNQ} Z_N [v] &= C_N \; \Pf \big[ \gamma_{j,k} [v] \big]_{j,k= 1, \dots ,2N},
\end{align}
where
\begin{align*}
\gamma_{j,k} [v] &:= \frac1{i} \int_{\mathbb{C}_{+}} (dz)\; v(z) h (z) h (\cc{z}) \left( p_{j-1}(z) p_{k-1} \left(\cc{z} \right) - p_{j-1}\left(\cc{z} \right) p_{k-1} (z) \right)
\end{align*}
and the $p_j (z)$ are monic polynomials of degree $j$.
\end{proposition}

Note that
the choice of
the polynomials $p_{j} (x)$ is not unique;
indeed, following through the construction of the Pfaffian generalized partition function we find that we may choose any polynomials that satisfy 
the
criteria
of being monic and of degree $j$.
So, the task of obtaining the correlation functions will be greatly simplified if the polynomials can be chosen such that they \textit{skew-orthogonalize} the matrix in (\ref{e:ZNQ}), that is they reduce it to the form of \eqref{e:skew_diag_mat}, where the diagonal blocks are the $2\times 2$ matrices
\begin{align}
\nonumber \left[ \begin{array}{cc}
0 & g_j\\
-g_j & 0
\end{array}\right],
\end{align}
with $g_j=\gamma_{2j+1, 2j+2}[1]$. Specifically, we define the skew-symmetric inner product
\begin{align}
\label{e:soips} \langle p_j, p_k \rangle := \gamma_{j+1,k+1}[1],
\end{align}
and look for polynomials to satisfy the skew-orthogonality conditions
\begin{align}
\label{e:soc} \langle p_{2j}, p_{2k} \rangle = \langle p_{2j+1}, p_{2k+1} \rangle =0 &, &\langle p_{2j}, p_{2k+1} \rangle = -\langle p_{2k+1} ,p_{2j} \rangle= \delta_{j,k} \: g_j.
\end{align}
We call these \textit{skew-orthogonal polynomials}. Assuming the existence of polynomials satisfying (\ref{e:soc}) (these polynomials do indeed exist, see \eqref{e:SOPS4}) then we can follow \cite{ForrNaga2007, BoroSinc2009, Forr2010} to calculate the eigenvalue correlation functions from the generalized partition function above. We use the identity $\det(1+ AB)= \det (1+ BA)$ for general linear operators, or a Pfaffian or quaternion determinant analogue, to write the generalized partition function (\ref{e:ZNQ}) as a Fredholm Pfaffian or quaternion determinant (see Appendix \ref{app:quatPf}),
\begin{align}
\nonumber \Pf[1+\lambda K] &:=1+\sum_{s=1}^{\infty}\frac{\lambda^s}{s!} \int_{-\infty}^{\infty} dx_1 \cdots \int_{-\infty}^{\infty} dx_s\; \Pf [\mathbf{K}(x_j, x_k)]_{j,k=1, \dots ,s}\;,\\
\nonumber \qdet[1+\lambda \tilde{K}] &:=1+\sum_{s=1}^{\infty}\frac{\lambda^s}{s!} \int_{-\infty}^{\infty} dx_1 \cdots \int_{-\infty}^{\infty} dx_s\; \qdet [\tilde{\mathbf{K}}(x_j, x_k)]_{j,k=1, \dots ,s}\;,
\end{align}
which can then be substituted into (\ref{e:correlnd}) to immediately yield the correlation functions, with Pfaffian kernels
\begin{align}
\nonumber \bK (x, y) :=\left[
\begin{array}{cc}
 D (x, y) & S (x, y) \\
 -S (y, x) &  I (x, y)
\end{array}
\right].
\end{align}
The details of the calculation are by now well established, and somewhat involved, so we refer the reader to the papers mentioned above, as well as to \cite{Forr2010, Mays2011}.
\begin{proposition}\label{p:correlns4}
With polynomials $q_1, q_2 \dots$ skew-orthogonal with respect to the inner product $\langle p_j, p_k \rangle$ of \eqref{e:soips} the $m$-point correlation function for the real quaternion induced spherical ensemble is
\begin{align}
\label{e:correlnsQ} \rho_{(m)} (\mathbf{z}_m)=\Pf\left[
\bK (z_s, z_t) \right]_{s,t=1, \dots , m} ,\qquad z_i\in \mathbb{C}_+,
\end{align}
where
\begin{align}
\nonumber D (x, y)&= i \sum_{j=0}^{N -1} \frac{h (x) h (y)} {g_j} \left( q_{2j} (x) q_{2j+1}(y)- q_{2j+1}(x) q_{2j}(y) \right),\\
\nonumber S (x, y) &= i \sum_{j=0}^{N -1} \frac{h (x) h (\cc{y})} {g_j} \left( q_{2j}(x) q_{2j+1} (\cc{y})- q_{2j+1}(x) q_{2j} (\cc{y}) \right),\\
\nonumber I (x, y) &= i \sum_{j=0}^{N -1} \frac{ h(\cc{x}) h (\cc{y})} {g_j} \left( q_{2j} (\cc{x}) q_{2j+1}(\cc{y}) -q_{2j+1}(\cc{x}) q_{2j} (\cc{y}) \right).
\end{align}
\end{proposition}

Note that
\begin{align}
\label{e:kerrelns4} D (x, y) = S (x, \cc{y}) = I (\cc{x}, \cc{y}).
\end{align}

\section{Skew-orthogonal polynomials}
\label{s:AvcharSOPS}

The expressions for the correlation kernel elements $S$, $D$, $I$ given in Proposition \ref{p:correlns4} depend on the skew-orthogonal polynomials (that is, polynomials satisfying \eqref{e:soc}) --- once we have these polynomials, then we have full knowledge of the correlation functions. In previous studies good use has been made of averages over characteristic polynomials to access the skew-orthogonal polynomials, or to avoid them entirely (see for example \cite{BaikDeifStra2003, FyodSomm2003, Akem2005, FyodKhor2007a, AkemPhilSomm2009, KhorSommZycz2010, KhorSomm2011, Forr2010a}). In particular, for the real analogue of the real quaternion ensemble of this paper, \cite{Fisc2013} uses exactly this method to find the skew-orthogonal polynomials and the eigenvalue density corresponding to \eqref{e:dens4}.

However, the situation is somewhat different in the real quaternion case that we consider here: while we are able to write down an expression for the average over the characteristic polynomial in terms of the skew-orthogonal polynomials analogous to \cite[Corollary 4.1.11]{Fisc2013}, it is not known how to evaluate it. In the following proposition we state this expression using a method of proof similar to that in \cite{ForrMays2011}.
\begin{proposition}
With the characteristic polynomial for a real quaternion matrix,
\begin{align}
\nonumber \phi(x) := \prod_{j=1}^N (x- \lambda_j) (x- \cc{\lambda}_j),
\end{align}
we have
\begin{align}
\label{e:chavQ1} S (z, w) = \frac{C_{N}} {i^{N-2} C_{N-1}} \ (\overline{w} - z) h (z) h (\overline{w}) \left\langle \phi(z) \phi(\overline{w}) \right\rangle_{\mathcal{Q} \big|_{N\mapsto N-1}},
\end{align}
where the average is over the jpdf (\ref{e:ejpdfQ}) with $N-1$ eigenvalues.
\end{proposition}

\proof We start by writing
\begin{align}
\nonumber &\phi(z) \phi(\overline{w}) \mathcal{Q} (\boldsymbol{\lambda}_N, \boldsymbol{\cc{\lambda}}_N ) = \frac{C_{N}} {\Gamma(N+1)(\cc{w}- z)} \det \left[ \begin{array}{c}
\left[ \begin{array}{c}
h(\lambda_j) p_{k-1}(\lambda_j)\\
h(\cc{\lambda_j}) p_{k-1}(\cc{\lambda}_j)
\end{array} \right]_{j=1, \dots, N}\\
p_{k-1} (z)\\
p_{k-1} (\cc{w})
\end{array} \right]_{k= 1, \dots, 2N+2},
\end{align}
using the Vandermonde identity.

Integrating over the independent elements of the eigenvalues we obtain the ensemble average with respect to the density $\mathcal{Q}$,
\begin{align}
\label{e:chavQ2} \langle \phi(z) \phi(\overline{w}) \rangle_{\mathcal{Q}}= \frac{i^N C_{N}} {\overline{w} - z} \left[ \varkappa^N \right] \Pf \left[ \varkappa \gamma_{j, k} [1]+ \sigma_{j,k} \right]_{\genfrac{}{}{0pt}{}{j=1, \dots, 2N+2}{k= 1,\dots, 2N+2}},
\end{align}
where $[\varkappa^N]$ denotes that we take the coefficient of $\varkappa^N$, and
\begin{align}
\nonumber \sigma_{j,k}= p_{j-1}(z) p_{k-1}(\overline{w})- p_{j-1}(\overline{w}) p_{k-1}(z).
\end{align}
Using the fact that $Z_{N+1} [1]=1=C_{N+1}\prod_{k=1}^{N+1} \gamma_{2k-1, 2k} [1]$ in \eqref{e:ZNQ}, then with the polynomials $p_j$ equal to the skew-orthogonal polynomials $q_j$, we expand the Pfaffian on the RHS of \eqref{e:chavQ2} and obtain
\begin{align}
\nonumber \langle \phi(z) \phi(\overline{w}) \rangle_{\mathcal{Q}}= \frac{i^N C_{N}} {(\overline{w}- z) C_{N+1}} \sum_{j=1}^{N+1} \frac{\sigma_{2j-1, 2j}}{\gamma_{2j-1, 2j} [1]} = \frac{i^{N-1}C_{N}} {C_{N+1}} \frac{(\overline{w}- z)^{-1}} {h(z) h(\overline{w})} \left( D (z, \overline{w})\big|_{N\mapsto N+1}\right).
\end{align}
Noting the relation between $S$ and $D$ in (\ref{e:kerrelns4}) we have the result on relabeling $N\mapsto N-1$.

\hfill $\Box$

In principle, one can use \eqref{e:chavQ1} to find the polynomials that skew-orthogonalize the Pfaffian in \eqref{e:ZNQ}, but fortunately the required polynomials have been already found.

\begin{proposition}[\cite{Forr2013}, Proposition 4]
The polynomials that skew-orthogonalize the inner product $\nonumber \langle p_j, p_k \rangle$ are
\begin{align}
\label{e:SOPS4} q_{2k+1}(z)= z^{2k+1},&& q_{2k}(z)&= \frac{\Gamma(L+k+1)} {\Gamma (k+1/2-n)} \sum_{j=0}^k (-1)^{k-j} \frac{\Gamma (j+1/2-n)} {\Gamma (j+1+L)} z^{2j},
\end{align}
which gives the normalization
\begin{align}
\nonumber g_k &:= \gamma_{2k+1, 2k+2}[1]= \pi \frac{\Gamma (2n-2k) \Gamma(2L+2k+2)} {\Gamma (2n+2L+2)}, \quad k=0, \dots ,N-1.
\end{align}
\end{proposition}

With these polynomials, and the relations in (\ref{e:kerrelns4}) we have fully specified the Pfaffian kernel in Proposition \ref{p:correlns4}, that is, substituting the polynomials into $S (z, w)$ we have
\begin{align}
\nonumber S (z,w)&= \frac{\Gamma (2n+2L+2)}{i \:2^{2(L+n)}} \frac{|z|^{2L} (z-\overline{z})^{1/2}} {(1+|z|^2)^{n+L+1}} \frac{|w|^{2L} (\overline{w}-w)^{1/2}} {(1+|w|^2)^{n+L+1}}\\
\label{e:szw2b} &\times \sum_{k=0}^{N-1} \sum_{j=0}^k \frac{ \left(z^{2j} \cc{w}^{2k+1} - z^{2k+1} \cc{w}^{2j} \right)} {\Gamma (L+j+1) \Gamma (L+k+3/2) \Gamma (n-j+1/2) \Gamma (n-k)}.
\end{align}
The density is given by $\rho_{(1)} (z) = S (z, z)$, which gives us (\ref{e:dens4}). Although it is not known how to write this sum in closed form, in Section \ref{s:slims} we are able to make use of an integral approximation for large $N$ to obtain asymptotic results. In Section \ref{s:gencorrelns4}, we propose a differential equation, which, if it could be solved, would give a closed form expression for \eqref{e:szw2b} as was done in \cite{Kanz2002, Akem2005, Mays2013}.

\section{Limiting densities}
\label{s:slims}

\begin{figure}[!htb]
\begin{center}
\subfloat[]{\includegraphics[scale=0.7]{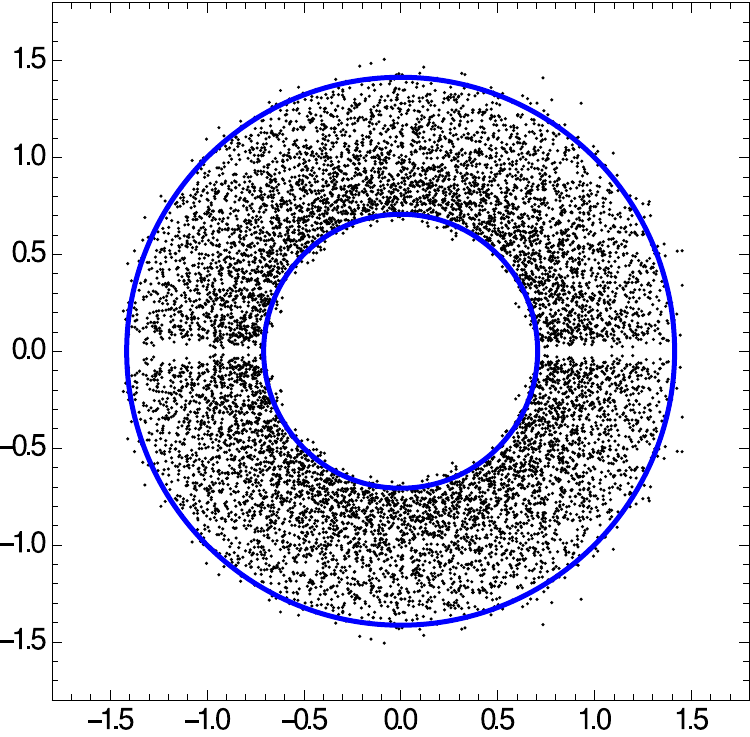}} \qquad\qquad
\subfloat[]{\includegraphics[scale=0.4]{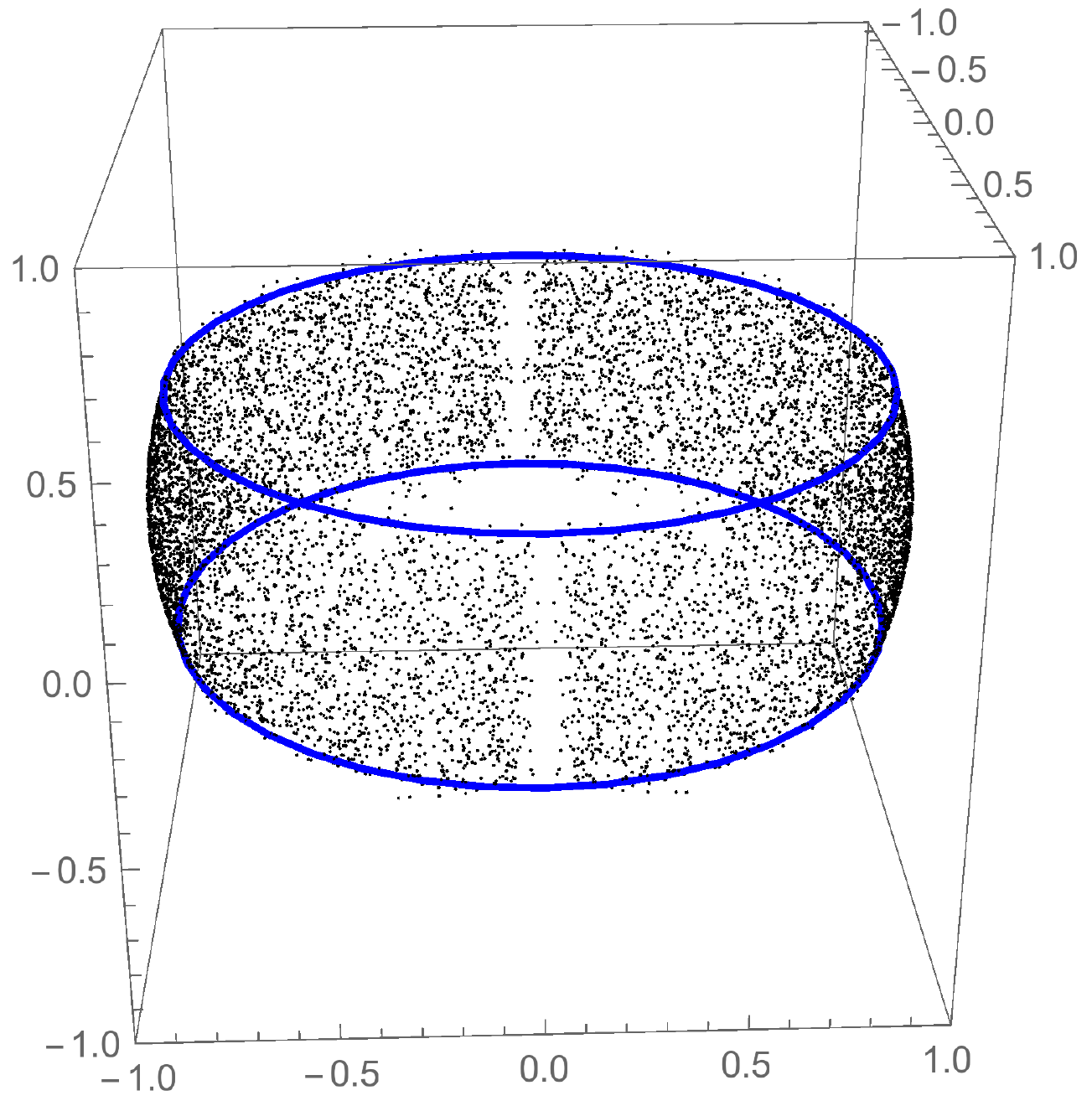}}
\end{center}
\caption{Eigenvalue plots for $\beta=4$, $N=100$, $L=100$, $n=200$ and $50$ realizations, (a) on the plane, and (b) after stereographic projection to the unit sphere. The blue lines indicate the radii $r_{\mathrm{in}}$ and $r_{\mathrm{out}}$.}
\label{f:quatBB}
\end{figure}

\begin{figure}[!bht]
\begin{center}
\subfloat[]{\includegraphics[scale=0.7]{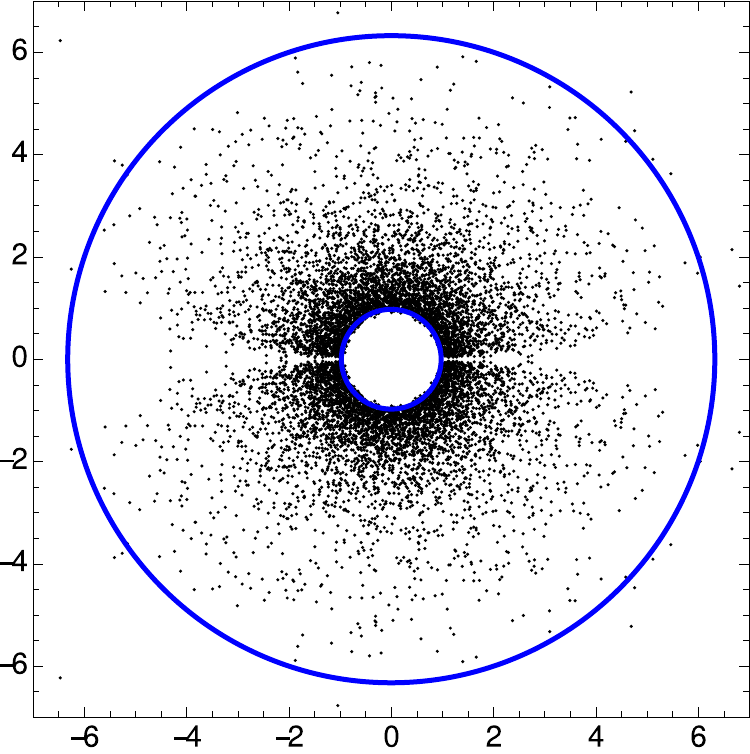}} \qquad\qquad
\subfloat[]{\includegraphics[scale=0.4]{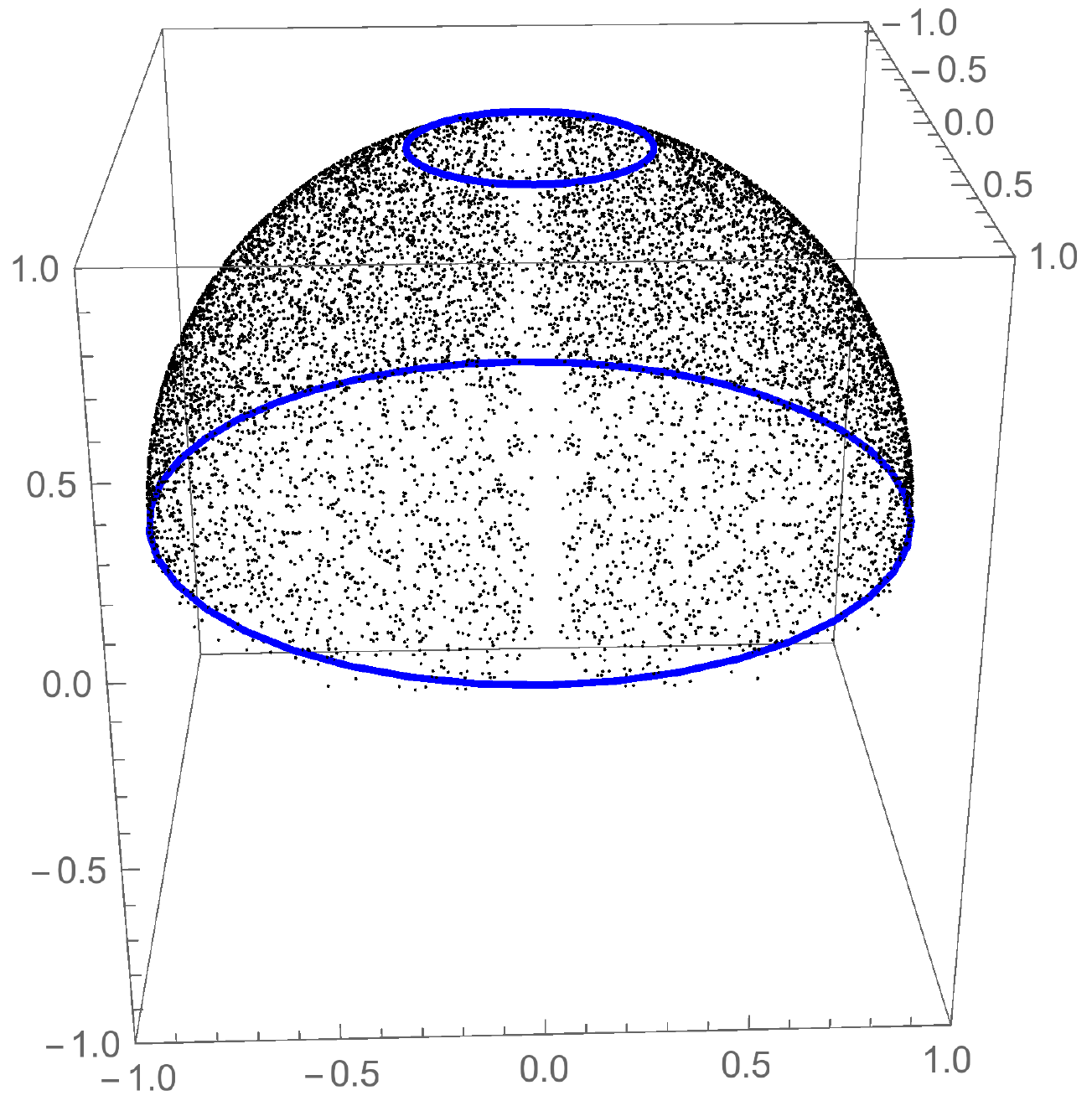}}
\end{center}
\caption{Eigenvalue plots for $\beta=4$, $N=100$, $L=100$, $n=105$ and $50$ realizations, (a) on the plane, and (b) after stereographic projection to the unit sphere. The blue lines indicate the radii $r_{\mathrm{in}}$ and $r_{\mathrm{out}}$.}
\label{f:quatSB}
\end{figure}

\begin{figure}[!bht]
\begin{center}
\subfloat[]{\includegraphics[scale=0.7]{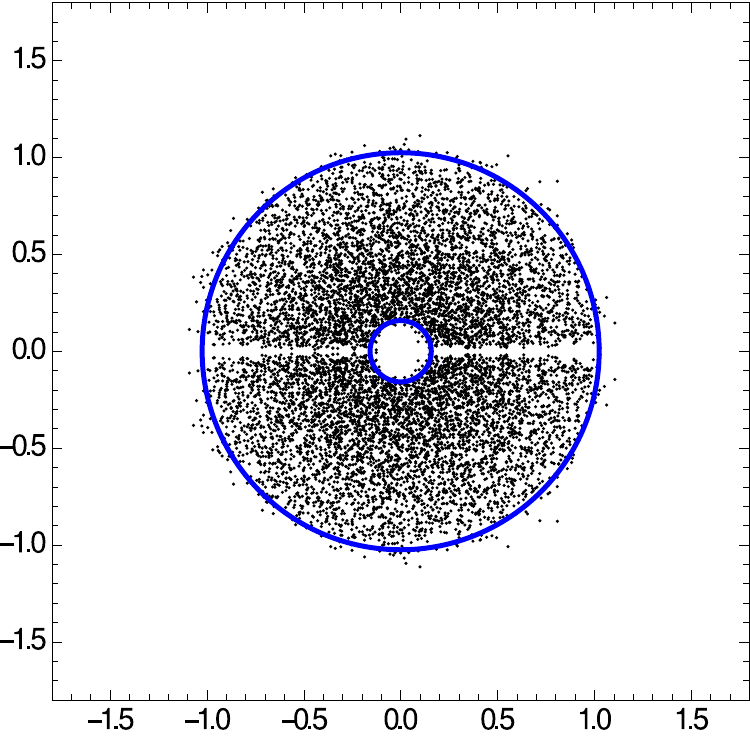}} \qquad\qquad
\subfloat[]{\includegraphics[scale=0.4]{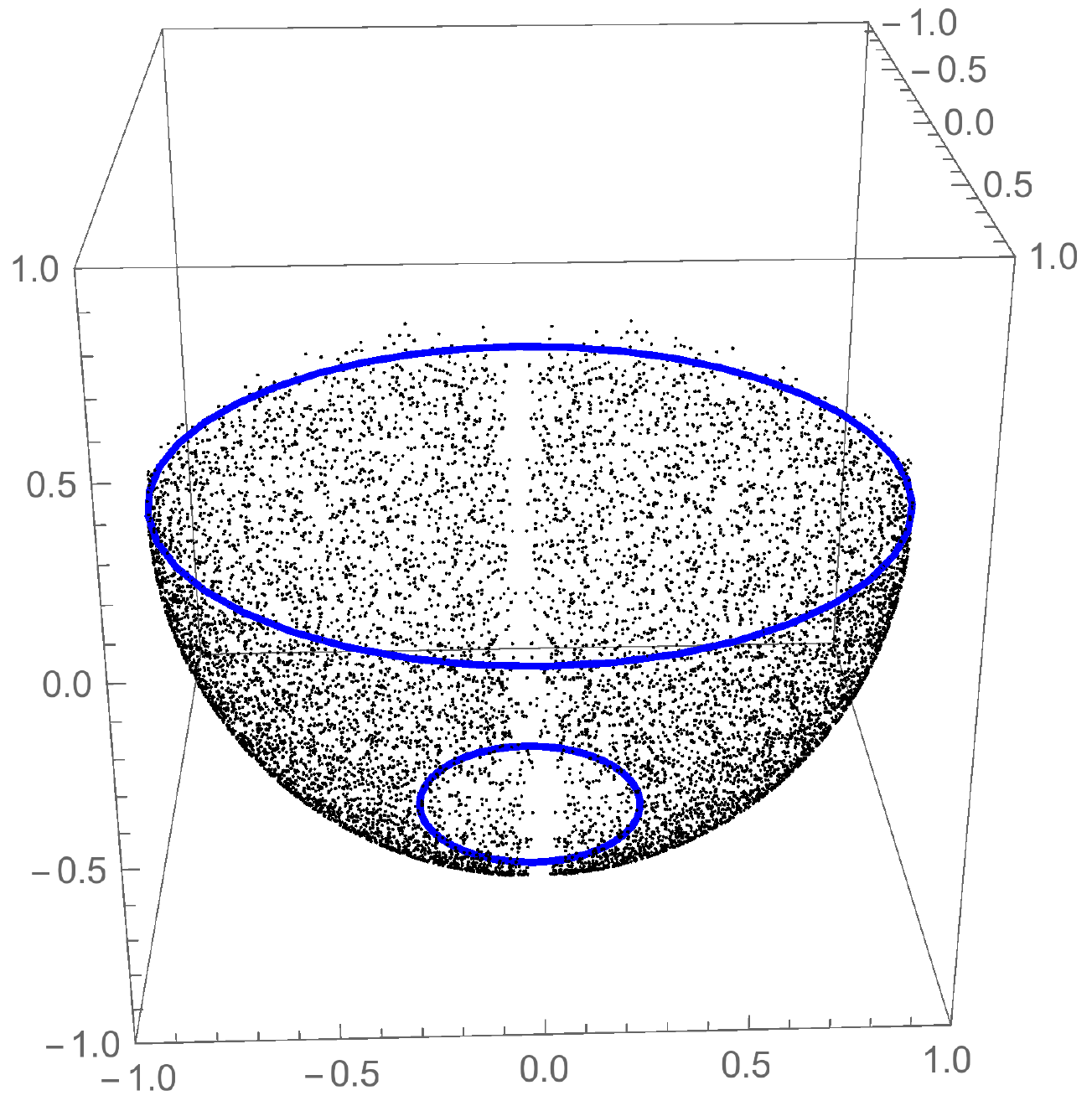}}
\end{center}
\caption{Eigenvalue plots for $\beta=4$, $N=100$, $L=5$, $n=200$ and $50$ realizations, (a) on the plane, and (b) after stereographic projection to the unit sphere. The blue lines indicate the radii $r_{\mathrm{in}}$ and $r_{\mathrm{out}}$.}
\label{f:quatBS}
\end{figure}

\begin{figure}[!bht]
\begin{center}
\subfloat[]{\includegraphics[scale=0.7]{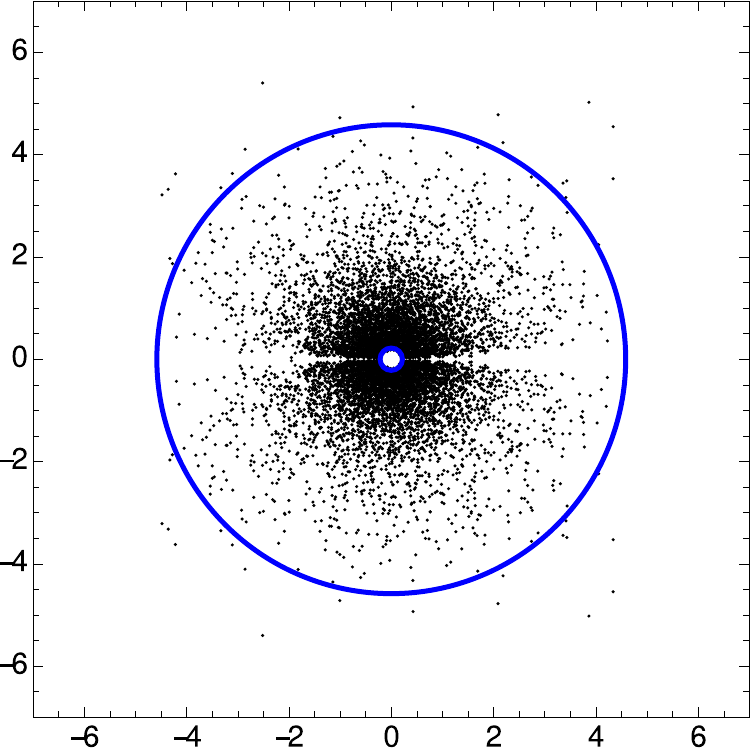}} \qquad\qquad
\subfloat[]{\includegraphics[scale=0.4]{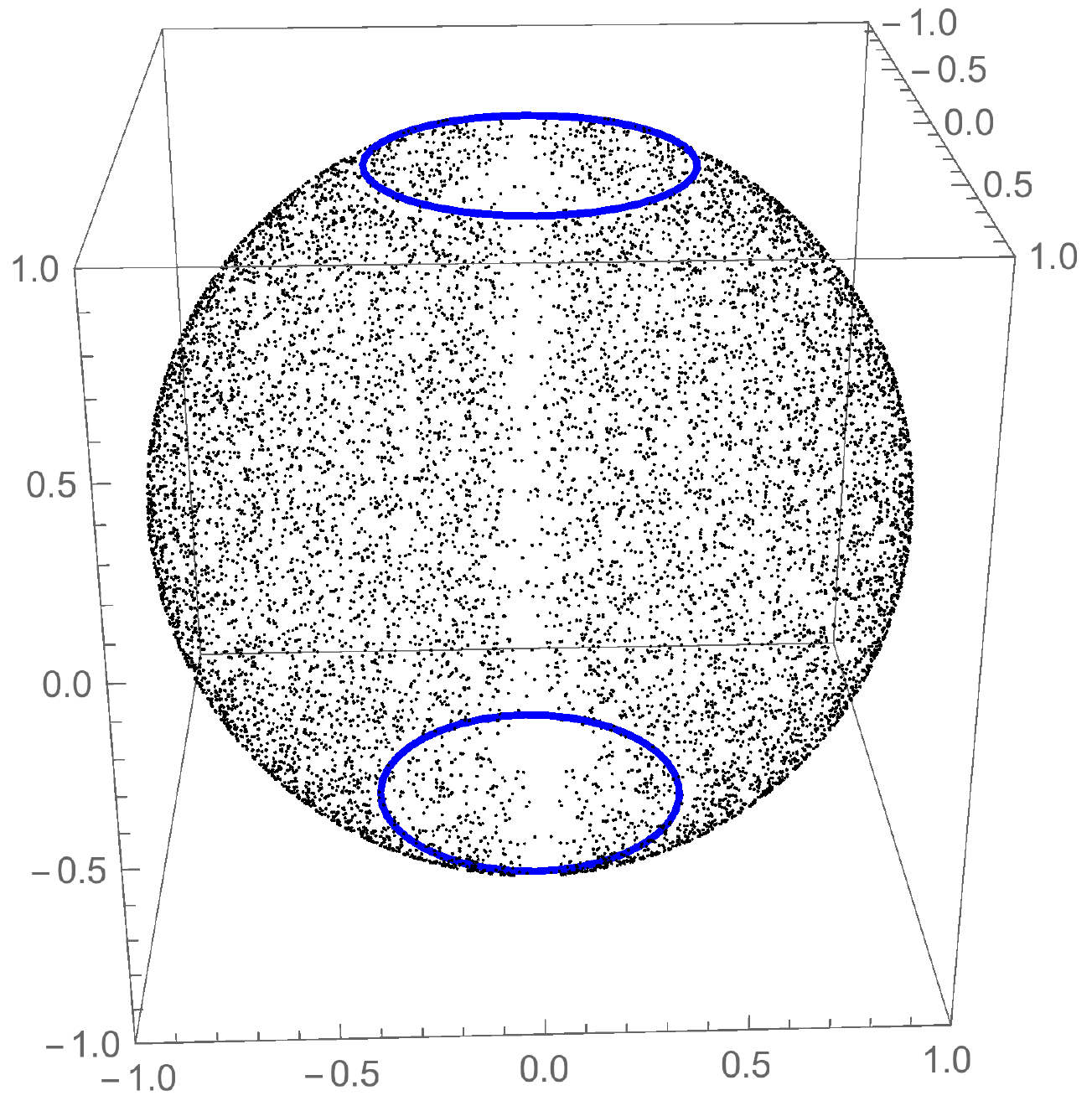}}
\end{center}
\caption{Eigenvalue plots for $\beta=4$, $N=100$, $L=5$, $n=105$ and $50$ realizations, (a) on the plane, and (b) after stereographic projection to the unit sphere. The blue lines indicate the radii $r_{\mathrm{in}}$ and $r_{\mathrm{out}}$.}
\label{f:quatSS}
\end{figure}

As discussed in the Introduction, the unique status of the real line distinguishes the eigenvalue density in the real ($\beta=1$), complex ($\beta=2$) and real quaternion ($\beta=4$) non-Hermitian ensembles; as such there are various universality results relating the eigenvalue density for the three classes of non-Hermitian ensembles away from the real axis. Based upon the asymptotic results of \cite{FiscForr2011, FBKSZ2012, Fisc2013} we can draw upon this concept of universality to expect that the limiting behaviour of the eigenvalue density (away from the real axis) falls into four regimes based on scaling of the parameters $n, L$:
\begin{align*}
\bullet\quad L&=aN, &\hspace{-3cm} n-N&= bN,\\
\bullet\quad L&=o(N), &\hspace{-3cm} n-N&= bN,\\
\bullet\quad L&=aN, &\hspace{-3cm} n-N&= o(N),\\
\bullet\quad L&=o(N), &\hspace{-3cm} n-N&= o(N),
\end{align*}
where $a,b$ are some constants. The classes are distinguished by the support of the limiting eigenvalue density, and we find (after inverse stereographic projection) that the bulk density for the real quaternion ensemble is uniform on a spherical annulus, conforming to the universality result known as the spherical law \cite{Bord2011}. The annulus for $\beta=4$ has the same inner and outer radii as in the $\beta=1$ and $\beta=2$ ensembles, namely
\begin{align}
\label{d:rinout} r_{in}^2:= \mu_1:= \frac{L}{n}, \qquad r_{out}^2:= \mu_2:= \frac{N+ L} {n- N}.
\end{align}
By way of illustration, we refer to Figures \ref{f:quatBB}--\ref{f:quatSS}, which display these regimes graphically. In the figures we have plotted the eigenvalues of $50$ independent $200\times 200$ random induced real quaternion spherical matrices in the complex plane, and then on the sphere (using inverse stereographic projection); one can see that as $L$ and $n- N$ increase, the eigenvalues tend to cluster closer to the equator of the sphere. (The generation of these matrices is described in Appendix \ref{s:genrand}.) The solid blue rings in the plots are $r_{in}$ and $r_{out}$. Note that the distinctive $\beta =4$ depletion of eigenvalues along the real line is clearly visible.

\subsection{Away from the real line}
The form of the double sum in the eigenvalue density \eqref{e:dens4} prevents us from using the differential equation methods of \cite{Kanz2002, Akem2005, Mays2013} to obtain asymptotic estimates of (\ref{e:dens4}) (see Section \ref{s:gencorrelns4} for more on this point). However, in this section we are focussing on the asymptotic results away from the real axis and in this region universality tells us that the eigenvalue density will be rotationally symmetric in the large $N$ limit. So we begin by integrating over the phase, which has the result of removing one of the sums,
\begin{align}
\nonumber r dr \int_{0}^{\pi} d\theta \; \rho_{(1)} (r e^{i \theta}) &=  \frac{2 r^{4L +3}}{(1+ r^2)^{2n+2L+2}} \sum_{k=0}^{N-1} \frac{\Gamma (2n+ 2L +2) r^{4k}} {\Gamma (2L+ 2k +2) \Gamma (2n -2k)} dr\\
\label{e:denstheta} &=: \bar{\rho}_{(1)} (r) dr.
\end{align}
To further aid the analysis, we let $N k'= k, \Delta k = N \Delta k'$ and replace the sum in \eqref{e:denstheta} with an integral approximation that will be accurate in the large $N$ limit:
\begin{align}
\label{e:sumint}\sum_{k=0}^{N-1} \frac{r^{4k}} {\Gamma (2L+ 2k +2) \Gamma (2n -2k)} \mathop{\rightarrow}\limits_{N \to \infty} N \int_0^1 \frac{r^{4N k'}} {\Gamma (2L+ 2N k' +2) \Gamma (2n -2N k')} dk'.
\end{align}
This integral approximation allows us to control the size of the arguments of the $\Gamma$ functions when applying Stirling's approximation (once we have specified the asymptotic behaviour of the parameters $n$ and $L$).

\subsubsection{Large $L$ and large $n-N$}
\label{s:srss}

This is the first of the asymptotic regimes mentioned above; here we let $L=aN$ and $n=(b+1)N$ for some constants $a,b$. We have tried to keep our notation consistent with \cite{FiscForr2011, FBKSZ2012, Fisc2013} to aid the reader.

\begin{proposition}\label{p:BB}
With $a,b$ some constants then let $L= aN$ and $n- N= bN$. In the limit of large matrix dimension $N$ the mean density of eigenvalues in the real quaternion induced spherical ensemble is
\begin{align}
\label{e:limdens4abulk} \lim_{N\to\infty} \frac{\rho_{(1)} (z)} {n +L} = \frac{2} {\pi} \frac{\Theta \left( |z| -r_{\mathrm{in}} \right) -\Theta \left( |z| -r_{\mathrm{out}} \right)} {(1+|z|^2)^{2} }.
\end{align}

At the edges of the annulus with radii $z_{\mathrm{in}} =\left( r_{\mathrm{in}}+ \frac{\xi} {\sqrt{n+L}} \right) \e^{i \phi}$ and $z_{\mathrm{out}}= \left( r_{\mathrm{out}} +\frac{\xi} {\sqrt{n+L}} \right) \e^{i \phi}$, we have
\begin{align}
\label{e:lim4in} \lim_{N \to \infty} \frac{\rho_{(1)} \left(z_{\mathrm{in}} \right)}{n+L} &= \frac{1} {\pi} \frac{1} {\left( 1+\mu_1 \right)^2}\erfc \left( \frac{-2 \xi} {1+ \mu_1} \right), \\
\label{e:lim4out} \lim_{N\to \infty} \frac{\rho_{(1)} \left(z_{\mathrm{out}} \right)} {n+L} &=\frac{1} {\pi} \frac{1} {\left( 1+\mu_2 \right)^2}\erfc \left( \frac{2 \xi} {1+ \mu_2} \right),
\end{align}
with $r_{in}, r_{out}, \mu_1, \mu_2$ as in \eqref{d:rinout}.

\end{proposition}

\proof

Making the replacements for $L, n$ in (\ref{e:denstheta}), and with \eqref{e:sumint}, we use Stirling's approximation to find the large $N$ behaviour of the product of gamma functions therein, giving us
\begin{align}
\nonumber \bar{\rho}_{(1)} (r) \sim \left( \frac{N} {\pi} \right)^{3/2} \frac{2 r^{4N \alpha +3} (\alpha + \beta +1)^{2N (\alpha +\beta +1) +3/2}}{(1+ r^2)^{2N (\alpha + \beta +1) +2}} \int_0^1 \frac{(\beta+1-k')^{1/2}} {(\alpha +k')^{3/2}} e^{N Y(k')} dk',
\end{align}
where
\begin{align}
\nonumber Y (k') = 4k' \log r - 2( \alpha +k') \log (\alpha +k') - 2(\beta+ 1- k') \log (\beta+ 1 -k').
\end{align}
We now have an expression suitable for the application of Laplace's method, which gives
\begin{align*}
\int_0^1 dk' \frac{(b+1-k')^{1/2}} {(a+k')^{3/2}} \exp[ N Y (k')] &\sim \frac{1} {2 r^{4 a N +2}}\left( \frac{1+r^2} {a+ b +1} \right)^{2N (a +b +1)} \sqrt{\frac{\pi } {N (a +b +1)}}\\
 & \times \erfc \left( \frac{a- (b+1)r^2} {r} \sqrt{\frac{N} {a +b +1}} \right).
\end{align*}
As part of this calculation, we have used the fact that $Y(k')$ is maximized at
\begin{align}
\nonumber k'_{max} &= \frac{(b+1) r^2- a} {1+r^2}\\
\nonumber \Rightarrow \quad \mu_1 &= \frac{a} {b +1} \leq r^2 \leq \frac{a +1} {b}= \mu_2,
\end{align}
since $k'_{max}\in [0,1]$. From the definition of the complementary error function we have
\begin{align}
\nonumber \erfc \left( \frac{a- (b+1)r^2} {r} \sqrt{\frac{N} {a +b +1}} \right) \mathop{\to}\limits_{N\to \infty} \left\{ \begin{array}{cc}
2, &\qquad r> r_{\mathrm{in}},\\
0, &\qquad r< r_{\mathrm{in}}.
\end{array}\right.
\end{align}
Putting these facts together we obtain (\ref{e:limdens4abulk}).

To obtain the inner and outer edge densities, we instead change variables
\begin{align}
\nonumber z_{\mathrm{in}} \rightarrow \left( r_{\mathrm{in}} + \frac{\xi} {\sqrt{N (a +b + 1)}} \right) e^{i \theta}
\end{align}
and we have
\begin{align}
\nonumber \erfc \left( \frac{a- (b+1)r^2} {r} \sqrt{\frac{N} {a +b +1}} \right) \to \erfc \left( \frac{-2 \xi} {1+ \mu_1} \right),
\end{align}
and similarly for $z_{\mathrm{out}}$.

\hfill $\Box$

So we have recovered the bulk result for the $\beta= 2$ induced spherical ensemble, while near the edges of the annulus we similarly recover the $\beta =2$ edge density \cite{FBKSZ2012, FiscForr2011}. Note that these results can also be directly related to those in the $\beta= 2$ Ginibre ensemble (see \cite{Gini1965, Forr2010}) by the rescaling $\xi\mapsto \sqrt{2} \xi/ (1+ r^2)$, where the meaning of $r$ depends on whether we are looking at the bulk or the edge.

\begin{figure}[!htb]
\begin{center}
\includegraphics[scale=0.7]{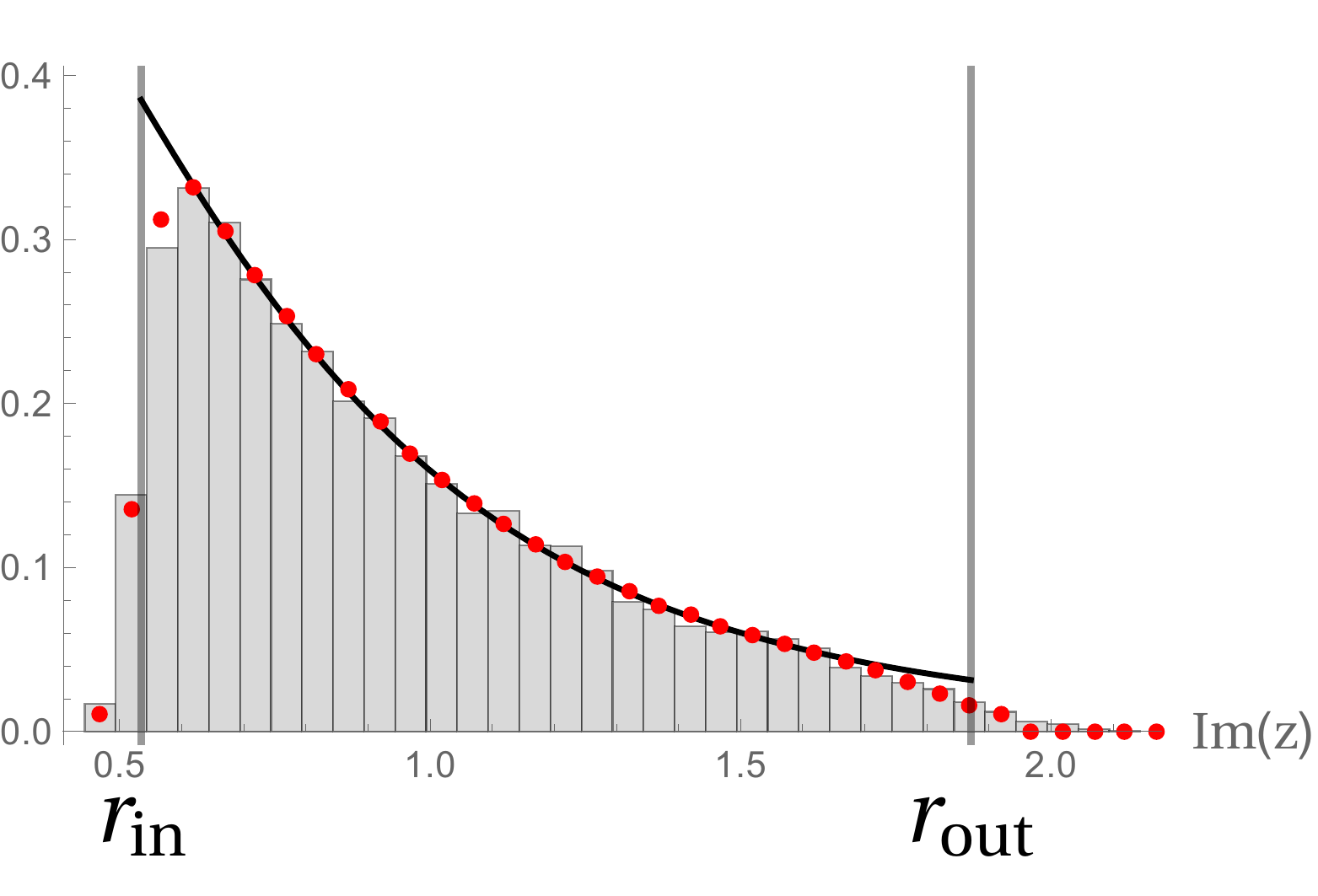}
\end{center}
\caption{A histogram of the empirical eigenvalue density along the imaginary axis with $N=100$, $n=140$, $L=40$ and $25,000$ realizations, taking those eigenvalues with $|Re(z)|<0.01$. The black line is the RHS of \eqref{e:limdens4abulk} (the limiting bulk density); the red dots are numerical evaluations of \eqref{e:dens4}, divided by $n+L$.}
\label{f:4bulkedens}
\end{figure}

\begin{figure}[!htb]
\begin{center}
\subfloat[]{\includegraphics[scale=0.42]{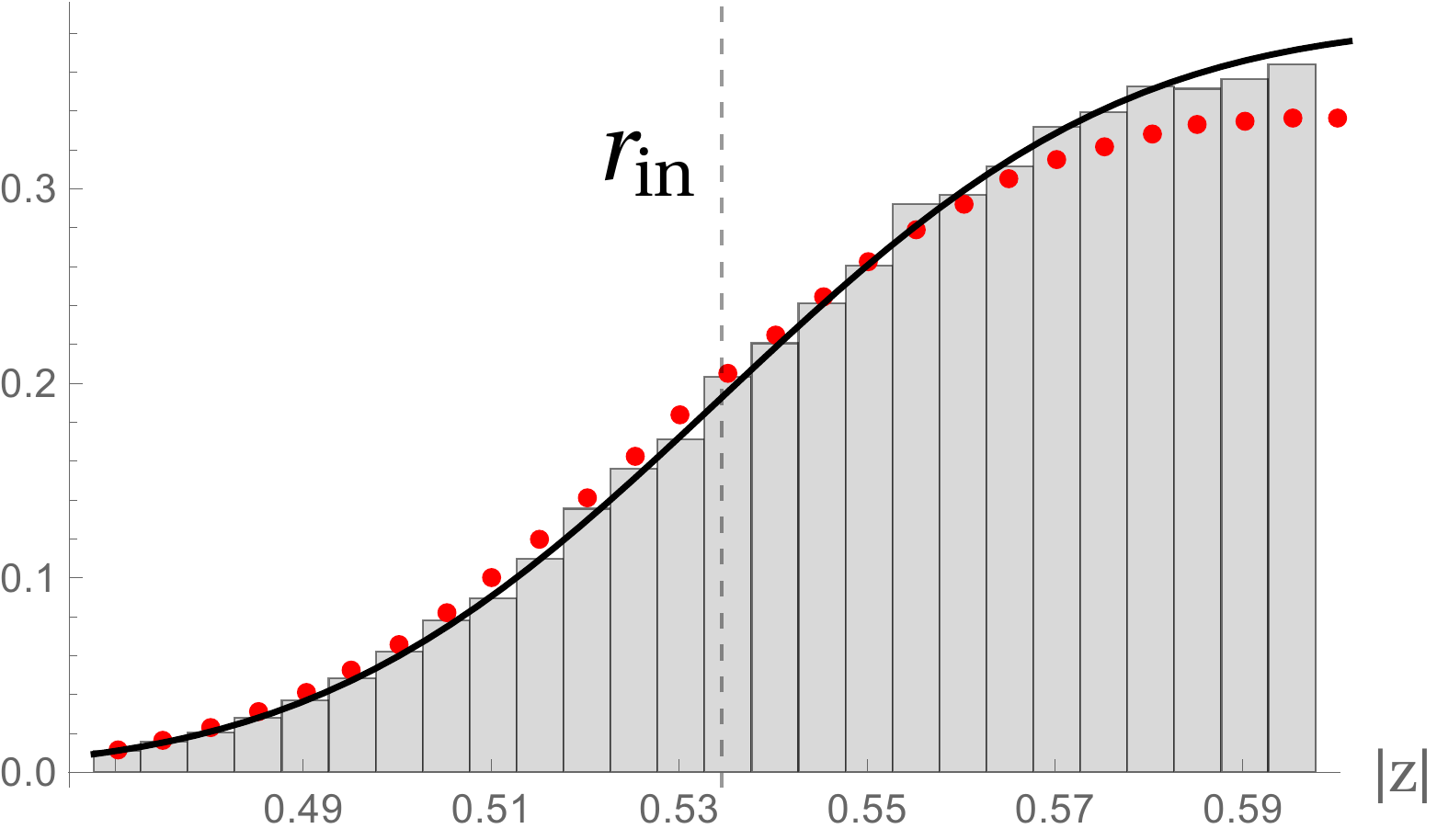}}\qquad
\subfloat[]{\includegraphics[scale=0.72]{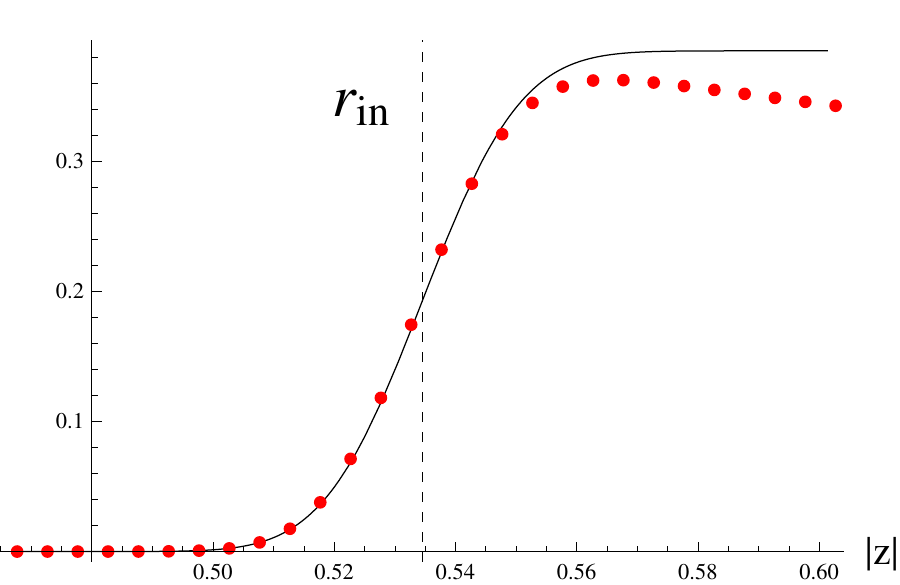}}
\end{center}
\caption{Density near the inner radius $r_{\mathrm{in}}$ for $\beta=4$. The solid black line is the limiting density \eqref{e:lim4in}; the red dots are numerical evaluations of \eqref{e:dens4}, divided by $n+L$. (a) Comparison to simulation with $N=100$, $n=140$, and $L=40$, taking those eigenvalues within a small distance of $r_{\mathrm{in}}$ and away from the real axis. (b) Demonstration that as the parameters increase (here $N=700$, $n=980$, and $L=280$), agreement between \eqref{e:lim4in} and \eqref{e:dens4} improves. One can see the transition to the bulk behaviour \eqref{e:limdens4abulk} as $r$ increases.}
\label{f:innerrad}
\end{figure}

We have simulated $25,000$ independent real quaternion induced spherical matrices (as described in Appendix \ref{s:genrand}) to compare with Proposition \ref{p:BB}. In Figure \ref{f:4bulkedens} we plot a histogram of the simulated eigenvalues taken from near the imaginary axis (where the repulsive effect of the real line is least felt), and compare it to the exact and bulk asymptotic results. In Figure \ref{f:innerrad}(a) we have made similar comparisons for the prediction of the eigenvalue density near the inner edge, while \ref{f:innerrad}(b) demonstrates that the agreement between the exact and asymptotic regimes improves as the parameters increase --- the transition to the bulk behaviour of Figure \ref{f:4bulkedens} can also be seen. We have provided a schematic illustration of the sampling regions in Figure \ref{f:sampreg}.


\subsubsection{Other scaling regimes}

In the other three regimes discussed at the beginning of Section \ref{s:slims}, $L$ or $n-N$ are kept small relative to $N$, and the limiting eigenvalue density annulus expands (see Figures \ref{f:quatSB}, \ref{f:quatBS} and \ref{f:quatSS}). The same reasoning as in Proposition \ref{p:BB} leads to the following modifications to \eqref{e:limdens4abulk}:
\begin{align*}
L&=o(N),&\!\!\! n-N &=bN: & r_{\mathrm{in}} &\mathop{\longrightarrow}\limits_{N\to \infty} 0&& \Rightarrow &\lim_{N\to\infty} \frac{\rho_{(1)} (z)} {n +L} &= \frac{2} {\pi} \frac{\Theta \left( r_{\mathrm{out}} -|z| \right)} {(1+|z|^2)^{2} },\\[4mm]
L&=aN,&\!\!\! n-N &=o(N): & r_{\mathrm{out}} &\mathop{\longrightarrow}\limits_{N\to \infty} \infty& & \Rightarrow &\lim_{N\to\infty} \frac{\rho_{(1)} (z) } {n +L} &= \frac{2} {\pi} \frac{\Theta \left( |z|-r_{\mathrm{in}} \right)} {(1+|z|^2)^{2} },\\[4mm]
L&=o(N),&\!\!\! n-N& =o(N): & r_{\mathrm{in}} & \mathop{\longrightarrow}\limits_{N\to \infty} 0 , &\!\!\! r_{\mathrm{out}} &\mathop{\longrightarrow}\limits_{N\to \infty} \infty \\
 &&&&&&& \Rightarrow &\lim_{N\to\infty} \frac{\rho_{(1)} (z)} {n +L} &= \frac{2} {\pi} \frac{1} {(1+|z|^2)^{2} }.
\end{align*}

Note that in the last case, when $L$ and $n-N$ are both small, we are in the regime of the (non-induced) spherical ensemble and we recover the result of \cite{Mays2013}.

In the cases that there remains an inner or outer edge, we retain \eqref{e:lim4in} and \eqref{e:lim4out}:
\begin{align*}
L&=o(N),& n-N &=bN: & \lim_{N\to \infty} \frac{\rho_{(1)} \left(z_{\mathrm{out}} \right)} {n+L} &=\frac{1} {\pi} \frac{1} {\left( 1+\mu_2 \right)^2}\,\erfc \left( \frac{2 \xi} {1+ \mu_2} \right),\\
L&=aN,& n-N &=o(N): & \lim_{N \to \infty} \frac{\rho_{(1)} \left(z_{\mathrm{in}} \right)}{n+L} &= \frac{1} {\pi} \frac{1} {\left( 1+\mu_1 \right)^2}\,\erfc \left( \frac{-2 \xi} {1+ \mu_1} \right).
\end{align*}

\subsection{Density near real line}
\label{s:gencorrelns4}

\begin{figure}[!bht]
\begin{center}
\includegraphics[scale=0.8]{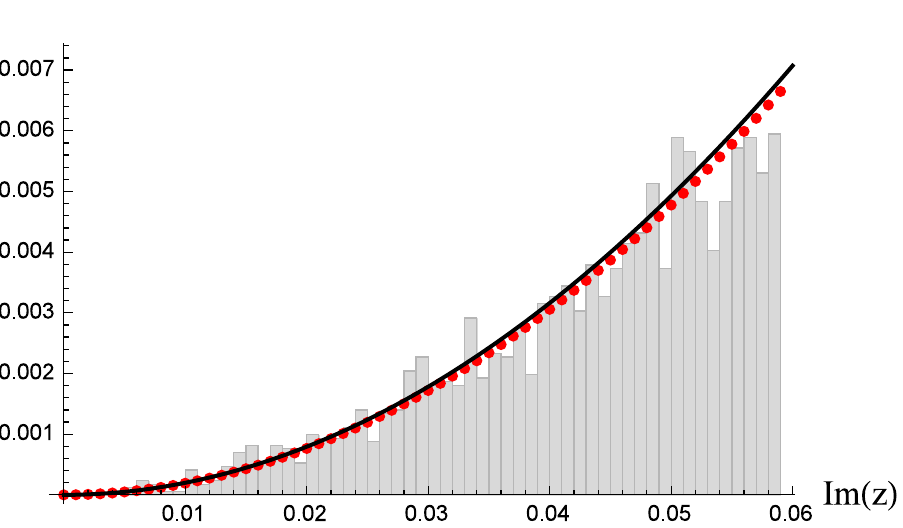}
\end{center}
\caption{A histogram of eigenvalues from a simulation with $N=100$, $n=140$, $L=40$, and $25,000$ realizations; shown along with a numerical evaluation of \eqref{e:dens4} divided by $\sqrt{n+L}$ (red dots) and a plot of \eqref{e:nearreal} with the constant chosen to be $1$ (black line).}
\label{f:nearrealevdens}
\end{figure}

With the rescaling of \cite{FiscForr2011, FBKSZ2012, Fisc2013}
\begin{align}
\nonumber r&= X+ \frac{x} {\sqrt{N (a + b +1)}}, \qquad \theta= \frac{y} {X \sqrt{N (a + b +1)}},
\end{align}
we should obtain the limiting density near the real line (the horizontal shift by $X$ is to ensure that we remain inside the annulus of support). By comparing the results for the real ensemble in \cite[Theorem 4.2.14]{Fisc2013} to those for the real Ginibre ensemble \cite{ForrNaga2007, BoroSinc2009} and the real spherical ensemble \cite{ForrMays2011}, we can conjecture the modifications needed to adapt the results from the real quaternion Ginibre ensemble \cite{Kanz2002} and spherical ensemble \cite{Mays2013} to the induced spherical ensemble:
\begin{align}
\label{e:nearreal} \frac{\rho (z_{real\; edge} )} {\sqrt{n +L}} \sim Const. \; \frac{4 \pi \; y\; e^{\frac{-8 \pi y^2}{(1+ x^2)^2}}} {i\; (1+ x^2)^3} \; \erfc \left( \frac{2 \sqrt{2 \pi}\; i\; y} {1+x^2} \right).
\end{align}

In Figure \ref{f:nearrealevdens} we compare the expression (\ref{e:nearreal}) to a simulated eigenvalue density near the real line (with $Const.=1$), which gives us confidence that our conjecture is correct, however we do not have any analytic results on this. As mentioned above, in previous studies on similar real quaternion ensembles \cite{Kanz2002, Akem2005} the calculation of analogous asymptotic results has been accomplished by finding a differential equation for the double sum equivalent to that in (\ref{e:dens4});\footnote{The calculation in \cite{Mays2013} was slightly different, due to the fractional linear transformation of the eigenvalues that was employed there.} solving this equation gives an expression for the double sum in terms of a single integral, which can then be analyzed asymptotically. This process relies on noting that the exponents of the variables are closely related to the factors appearing in the gamma functions, and so by taking derivatives one is able to reduce the size of the inner sum. A crucial part of this procedure is that a derivative with respect to (say) $z$ removes the terms proportional to $z^0$ from the sum. Then, by performing some judicious re-summing, one finds a soluble differential equation.

The problem we face here is that in order to obtain agreement between the exponent of $z$ and the gamma function factors, we must first multiply the sum through by $z^{2L}$, meaning that the first derivative with respect to $z$ will not kill off any terms in the sum, and so we obtain an equation containing iterated derivatives and anti-derivatives that has so far not been solved.

\begin{proposition}
Denote the double sum as
\begin{align}
\nonumber \sigma(z, w) := \sum_{k=0}^{N-1} \sum_{j=0}^k \frac{ \left(z^{2j} w^{2k+1} - z^{2k+1} w^{2j} \right)} {\Gamma (L+j+1) \Gamma (L+k+3/2) \Gamma (n-j+1/2) \Gamma (n-k)},
\end{align}
then we have the differential equations
\begin{align}
\nonumber \Big[ z^{2n} A_z z^{-2n- 2L -1} &(z A_z)^L D_z (z^{-1} D_z)^L z^{2L} + z \Big] \sigma (z, w)\\
\nonumber & = \frac{2^{2L +2n}} {\pi} \sum_{k=0}^{\infty} \frac{(z w)^L} {\Gamma (2L +k +1) \Gamma (2n -k +1)}\\
\label{e:dblsumDE} &=- \Big[ w^{2n} A_w w^{-2n- 2L -1} (w A_w)^L D_w (w^{-1} D_w)^L w^{2L} + w \Big] \sigma (z, w),
\end{align}
where $D_x = \frac{\partial} {\partial x}$ and $A_x$ is the anti-derivative (with constant term zero).

\end{proposition}

A similar problem has also been encountered in studying products of quaternionic Ginibre matrices, where equally intractable DEs were found \cite[(4.99) and (4.100)]{Ipse2015}.

One expects that the solution of the equation (\ref{e:dblsumDE}) will yield a `nice' integral expression for the eigenvalue correlation function kernel $S (x, y)$, then leading to asymptotic expressions for the full correlation function (\ref{e:correlnsQ}) in each of the four regimes of the parameters $L, n$. However, a solution to DEs like \eqref{e:dblsumDE} seems remote, and so it seems that different techniques will be required for the analysis of $\beta =4$ ensembles, since the double sum is a common feature (arising from the even skew-orthogonal polynomials \eqref{e:SOPS4}).

\section{Further work}

On the topic of universality results, we expect that the spherical law of \cite{Bord2011} can be generalized to the case here, a `spherical annulus law'
\begin{align}
\nonumber \frac{\rho_{(1)}^{(\beta)} (z)} {n +L} \mathop{\sim}\limits_{N \to \infty} \frac{\chi_{z\in S_A}} {\pi} ,
\end{align}
where $\chi_{\phi}$ is the indicator function, and $S_A$ is the spherical annulus corresponding to the boundary circles in the complex plane with radii \eqref{d:rinout}. We also suspect that the single-ring theorem of \cite{FeinZee1997} can be generalized to the case here, that is, the polynomial $V$ in \eqref{e:FZ} can perhaps be broadened to include the logarithmic expressions we find in \eqref{d:sphmats}.

Another outstanding calculation is that of the average over the product of characteristic polynomials
\begin{align*}
\langle \phi (z) \phi (\cc{w}) \rangle_{\mathcal{Q}}
\end{align*}
in \eqref{e:chavQ1}. Although the end result is known (by substitution of the skew-orthogonal polynomials \eqref{e:SOPS4}), it would be desirable to have a derivation along the lines of \cite[Theorem 4.2.9]{Fisc2013}, where the calculation reduced to an average over the orthogonal group, following the average over the unitary ensemble of \cite{FyodKhor2007a} --- a sense of symmetry compels one to feel that an average over the symplectic group will accomplish the task for $\beta =4$.

\section*{Acknowledgements}

The authors would particularly like to thank Jonith Fischmann for various assistance, including providing a copy of her thesis and some parts of code for simulating real induced spherical matrices. The authors would also like to thank Nicholas Beaton, Gaetan Borot, Peter Forrester, Yan Fyodorov, Jesper Ipsen, Boris Khoruzhenko, Francesco Mezzadri and Aris Moustakas for many thought-provoking discussions. They would also like to acknowledge the kind hospitality of: the School of Mathematics, University of Bristol; the School of Mathematical Science, Queen Mary University of London; the Zentrum f\"{u}r interdisziplin\"{a}re Forschung, Bielefeld Univsersity; and the Laboratoire de Physique Th\'{e}orique et Hautes Energies, University of Paris and Marie Curie. AP would like to acknowledge the support of the ERC grant 278124 ``LIC'', and the support of the Australian Research Council.

\appendix
\section{On quaternions, quaternion determinants and Pfaffians}
\label{app:quatPf}

Since quaternions are crucial to this study, we first provide a quick overview. A quaternion is analogous to a complex number, except that it has four basis elements instead of two. Typically they are written in the form
\begin{align}
\label{def:qns} q=q_0+iq_1+jq_2+kq_3,
\end{align}
with the relations $i^2=j^2=k^2=ijk=-1$, and the $q_l$ are in general complex. We will also use an alternative representation as $2\times 2$ matrices:
\begin{align}
\label{eqn:mat_quat} q=\left[ \begin{array}{cc}
w & x\\
y & z
\end{array}\right],
\end{align}
where $w=q_0+ iq_1, x=q_2+iq_3, y=-q_2+ iq_3, z=q_0- iq_1$. The analogue of complex conjugation for quaternions we denote $q^*=q_0-iq_1-jq_2-kq_3$, or in the matrix representation
\begin{align}
\nonumber q^*=\left[ \begin{array}{cc}
z & -x\\
-y & w
\end{array}\right],
\end{align}
and $|q|^2 =q_0^2 +q_1^2 +q_2^2 +q_3^2$. In the case that $q_0, q_1, q_2, q_3\in\mathbb{R}$ we say that $q\in \mathbb{H}$, the set of \textit{real quaternions}, and from (\ref{eqn:mat_quat}) we have
\begin{align}
\label{def:real_quats} q=\left[ \begin{array}{cc}
w & x\\
-\bar{x} & \bar{w}
\end{array}\right],
\end{align}
with conjugate
\begin{align}
\nonumber q^*=\left[ \begin{array}{cc}
\bar{w} & -x\\
\bar{x} & w
\end{array}\right].
\end{align}
In the $2\times 2$ representation, it is easy to see that $q^*=\det(q) q^{-1}=|q|^2 q^{-1}$, in analogy with complex numbers. With $\hat{\bQ}_{N\times N}= \left[ q_{j, k} \right]_{j,k= 1, \dots ,N}$ where $q_{j,k}\in \mathbb{H}$ (using the representation (\ref{def:qns})) we denote by $\hat{\bQ}_{N\times N}^D$ the matrix $[q^*_{k,j}]_{j,k= 1, \dots ,N}$, and we call it the \textit{dual} of $\hat{\bQ}_{N\times N}$. If $\hat{\bQ}_{N\times N}= \hat{\bQ}_{N\times N}^D$ then $\hat{\bQ}_{N\times N}$ is said to be \textit{self-dual}.

We will regularly use quaternion analogues of the usual matrix trace and determinant \cite{Dyso1970}.
\begin{definition}
\label{d:qqtd}
For an $N\times N$ matrix $\hat{\bQ}$ with real quaternion entries the \textit{quaternion trace} is defined as the sum of the scalar parts of the diagonal entries
\begin{align}
\label{d:qtr} \mathrm{qTr}\; \hat{\bQ} := \sum_{j=1}^N (q_0)_{j,j}.
\end{align}
The \textit{quaternion determinant} is defined by
\begin{equation}
\label{def:qdet} \qdet\; \hat{\bQ} :=\sum_{P\in S_N}(-1)^{N-|c(P)|}\prod_{(ab\cdots s)\in c(P)} \; \mathrm{qTr} (q_{ab}q_{bc}\cdots q_{sa}),
\end{equation}
where $c(P)$ is the set of cycles of the permutation $P$.
\end{definition}
Note that the definition (\ref{d:qtr})  gives
\begin{align}
\label{e:qtr} \mathrm{qTr}\; \hat{\bQ}_{N \times N} = \frac{1}{2} \Tr \; \bQ_{2N\times 2N},
\end{align}
where $\bQ_{2N\times 2N}$ is the matrix corresponding to $\hat{\bQ}_{N\times N}$ with the quaternions replaced by their $2\times 2$ representatives (\ref{def:real_quats}). Furthermore, it is shown in \cite{Dyso1970} that with the definition (\ref{def:qdet}) and with $\hat{\bQ}_{N\times N}$ a self-dual real quaternion matrix,
\begin{align}
\label{e:qdet} \qdet\; \hat{\bQ}_{N\times N} = \left( \det \bQ_{2N\times 2N} \right)^{1/2}.
\end{align}
Since we will be mostly using the $2\times 2$ representation for the quaternions we will most often make use of (\ref{e:qtr}) and (\ref{e:qdet}) instead of Definition \ref{d:qqtd}.

A structure that is closely related to the quaternion determinant is the Pfaffian.
\begin{definition}
\label{d:pf}
Let $\bX=[x_{ij}]_{i,j=1, \dots ,2N}$, where $x_{ji}=-x_{ij}$, so that $\bX$ is an anti-symmetric matrix of even size. Then the \textit{Pfaffian} of $\bX$ is defined by
\begin{align*}
\Pf [\bX]&=\sum^*_{P(2l)>P(2l-1)}\varepsilon (P) x_{P(1),P(2)}x_{P(3),P(4)}\cdots x_{P(2N-1),P(2N)}\\
 &=\frac{1}{2^NN!}\sum_{P\in S_{2N}}\varepsilon (P) x_{P(1),P(2)}x_{P(3),P(4)}\cdots x_{P(2N-1),P(2N)},
\end{align*}
where $S_{2N}$ is the group of permutations of $2N$ letters and $\varepsilon (P)$ is the sign of the permutation $P$. The * above the first sum indicates that the sum is over distinct terms only (that is, all permutations of the pairs of indices are regarded as identical).
\end{definition}

A classical result is that with $\bX$ as in Definition \ref{d:pf} we have
\begin{align}
\nonumber \left( \Pf\; \bX \right)^2 = \det \bX.
\end{align}
Usefully, Pfaffians can be calculated using a form of Laplace expansion. To calculate a determinant, recall that we can expand along any row or column. For example, expand a matrix $\bA=[a_{ij}]_{i,j=1, \dots n}$ along the first row:
\begin{equation}
\nonumber \det \bA =a_{1,1}\det[\bA]^{1,1}-a_{1,2}\det [\bA]^{1,2} + \cdots (-1)^{n+1} a_{1,n}\det [\bA]^{1,n},
\end{equation}
where $\det [\bA]^{i,j}$ means the determinant of the matrix left over after deleting the $i$th row and $j$th column.

The analogous expansion for a Pfaffian involves deleting \textit{two} rows and \textit{two} columns each time. For example, expanding a skew-symmetric matrix $\bB=[b_{ij}]_{i,j=1, \dots n}$ ($n$ even) along the first row:
\begin{equation*}
 \Pf\; \bB =b_{1,2}\Pf [\bB]^{1,2} -b_{1,3}\Pf [\bB]^{1,3} + \cdots + b_{1,n} \Pf[\bB]^{1,n},
\end{equation*}
where $\Pf[\bB]^{i,j}$ means the Pfaffian of the matrix left after deleting the $i$th  and $j$th rows and the $i$th and $j$th columns. Laplace expansion requires $n!$ calculations for a determinant, and $n!!=n\cdot (n-2)\cdot (n-4 )\cdot \dots $ in the case of a Pfaffian.

We recall that diagonal matrices have the property
\begin{align}
\nonumber \det\left( \mathrm{diag} [a_1, \dots ,a_N] \right) =\prod_{j=1}^N a_j,
\end{align}
and we can identify quaternion determinant and Pfaffian analogues of this statement. From (\ref{def:qdet}) we see that the analogous result for the quaternion determinant is
\begin{align*}
 \qdet\left( \mathrm{diag}[a_1,a_1, \dots ,a_{N/2}, a_{N/2}]\right) = \prod_{j=1}^{N/2} a_j.
\end{align*}
In the case of Pfaffians however, clearly diagonal matrices (with at least one non-zero element) are not skew-symmetric and so the Pfaffian of a diagonal matrix is undefined. However, we can define a suitably analogous matrix for a Pfaffian as
\begin{align}
\label{e:skew_diag_mat} \bT =\left[\begin{array}{cccc}
\bD_1 & \0 & \cdots & \0\\
\0 & \bD_2 & \cdots & \0\\
\vdots & \vdots & \ddots & \vdots\\
\0 & \0 & \cdots & \bD_{N/2}
\end{array}
\right],
\end{align}
where $\bD_j=\left[ \begin{array}{cc}
0 & a_j\\
-a_j & 0
\end{array}
\right]$ and $\0$ is the $2\times 2$ zero matrix. That is, the matrix has entries $\{a_1, \dots ,a_{N/2} \}$ along the diagonal above the main diagonal, and $\{-a_1, \dots ,-a_{N/2} \}$ on the diagonal just below the main diagonal, with zeros elsewhere. We call such a matrix \textit{skew-diagonal}, and
\begin{align}
\label{eqn:pf_skew_diag_eval} \Pf \; \bT =\prod_{j=1}^{N/2}a_j.
\end{align}
Note that in (\ref{e:skew_diag_mat}) and (\ref{eqn:pf_skew_diag_eval}), we have implicitly assumed that $N$ is even.

If we define
\begin{equation*}
 \bZ_{2N}:=\mathbf{1}_N\otimes \left[\begin{array}{cc}
0 & -1\\
1 & 0\\
\end{array}\right],
\end{equation*}
then for $\bM$ a $2N \times 2N$ self-dual matrix we have simple relations between the Pfaffian and quaternion determinant,
\begin{align}
\nonumber \Pf\; \bM\bZ^{-1}_{2N} &=\Pf\; \bZ^{-1}_{2N} \bM =\mathrm{qdet} \; \bM,\\
\nonumber \Pf\; \bM\bZ_{2N} &=\Pf \; \bZ_{2N} \bM =(-1)^N \mathrm{qdet} \; \bM.
\end{align}

\section{On the generation of random induced matrices}
\label{s:genrand}

In order to simulate matrices from the distribution of (\ref{d:sphmats}) we first define the \textit{rectangular spherical matrix}
\begin{align}
\label{e:rectYAB} \bY_{M\times N}^{\vphantom{-1/2}} = \bX_{M \times N}^{\vphantom{-1/2}} \bA^{-1/2}_{N \times N},
\end{align}
where $\bA \mathop{\sim}\limits^d W_N^{(\beta)} (n)$ is an $N\times N$ Wishart matrix with parameter $n$, and $\bX$ is an $M\times N$ Ginibre (iid) matrix with real, complex or real quaternion entries. Now we use the fact that the matrices \eqref{e:rectYAB} have the same distribution as the matrices \cite[Lemma 2.2.3]{Fisc2013}
\begin{align}
\label{d:Gtilde} \tilde{\bG}^{(\beta)}:= \bU^{(\beta)} (\bY^{\dagger} \bY)^{1/2},
\end{align}
where $\bY$ is from \eqref{e:rectYAB} and $\bU^{(\beta)}$ is a Haar distributed matrix which is: real orthogonal ($\beta=1$), complex unitary ($\beta= 2$), or symplectic (\textit{i.e.}, unitary real quaternion) ($\beta=4$).

The algorithm for generating the random induced spherical matrices relies on having a method to generate random Haar distributed matrices. In the real and complex case, this can be accomplished by applying the Gram--Schmidt algorithm to $N \times N$ random Gaussian matrices; (ignoring questions of numerical stability) a procedure that requires only a few lines of code in modern mathematical programming languages and takes $O(N^3)$ operations. However, quaternionic functionality is not as widely supported and so one needs to implement an algorithm from scratch. We implemented the algorithm described in \cite{Mezz2007}, which uses Householder transformations and also takes $O(N^3)$ operations, in addition to being more stable than Gram--Schmidt.\footnote{The algorithm for generating random symplectic matrices from \cite{Mezz2007} in this work was implemented using Octave (which is largely compatible with MATLAB). We are happy to share this code with the interested reader; it can be obtained by emailing AM.}

We used \eqref{d:Gtilde} to generate the eigenvalues in Figures \ref{f:quatBB}, \ref{f:quatSB}, \ref{f:quatBS} and \ref{f:quatSS}. We also used that construction to generate a set of $5,000,000$ eigenvalues from $25, 000$ independent matrices (with $N=100$, $n=140$, $L=40$) to obtain statistics with which to compare our expressions for the limiting eigenvalue densities in Section \ref{s:slims}. The first of these (Figure \ref{f:4bulkedens}) compares the finite $N$ density \eqref{e:dens4} and the bulk prediction \eqref{e:limdens4abulk} to the $16, 915$ eigenvalues with $\mathrm{Im} (z) >0, |\mathrm{Re} (z)|< 0.01$. We have chosen points near the imaginary axis, which should minimize distortions caused by the repulsion from the real line. 

Figure \ref{f:innerrad} compares the prediction \eqref{e:lim4in} for the density near the inner edge $r_{\mathrm{in}}\approx 0.5345$ of the annulus to the exact density \eqref{e:dens4} and the $100, 011$ eigenvalues from our simulation with $ 0.4677\approx r_{\mathrm{in}}- \frac{r_{\mathrm{out}}-r_{\mathrm{in}}}{20}< |z|< r_{\mathrm{in}}+\frac{r_{\mathrm{out}}-r_{\mathrm{in}}}{20}\approx 0.6013$, and $\pi/4< \mathrm{arg}(z)< 3\pi/4$. Again we have tried to maintain a balance between keeping a large number of eigenvalues, while discarding those close to the real line. Lastly, Figure \ref{f:nearrealevdens} again plots the exact density \eqref{e:dens4}, this time against the prediction \eqref{e:nearreal} (with the constant equal to $1$) for the density near the real line and the $2, 301$ eigenvalues with  $0 <\mathrm{Im} (z) < 0.06$ and $0.9354\approx \frac{r_{\mathrm{in}}+ r_{\mathrm{out}}} {2} -\epsilon_x < |Re (z)|< \frac{r_{\mathrm{in}}+ r_{\mathrm{out}}}{2} + \epsilon_x \approx 1.470$, where $\epsilon_x := \frac{r_{\mathrm{out}}-r_{\mathrm{in}}}{5}$.

\begin{figure}
\begin{center}
\includegraphics{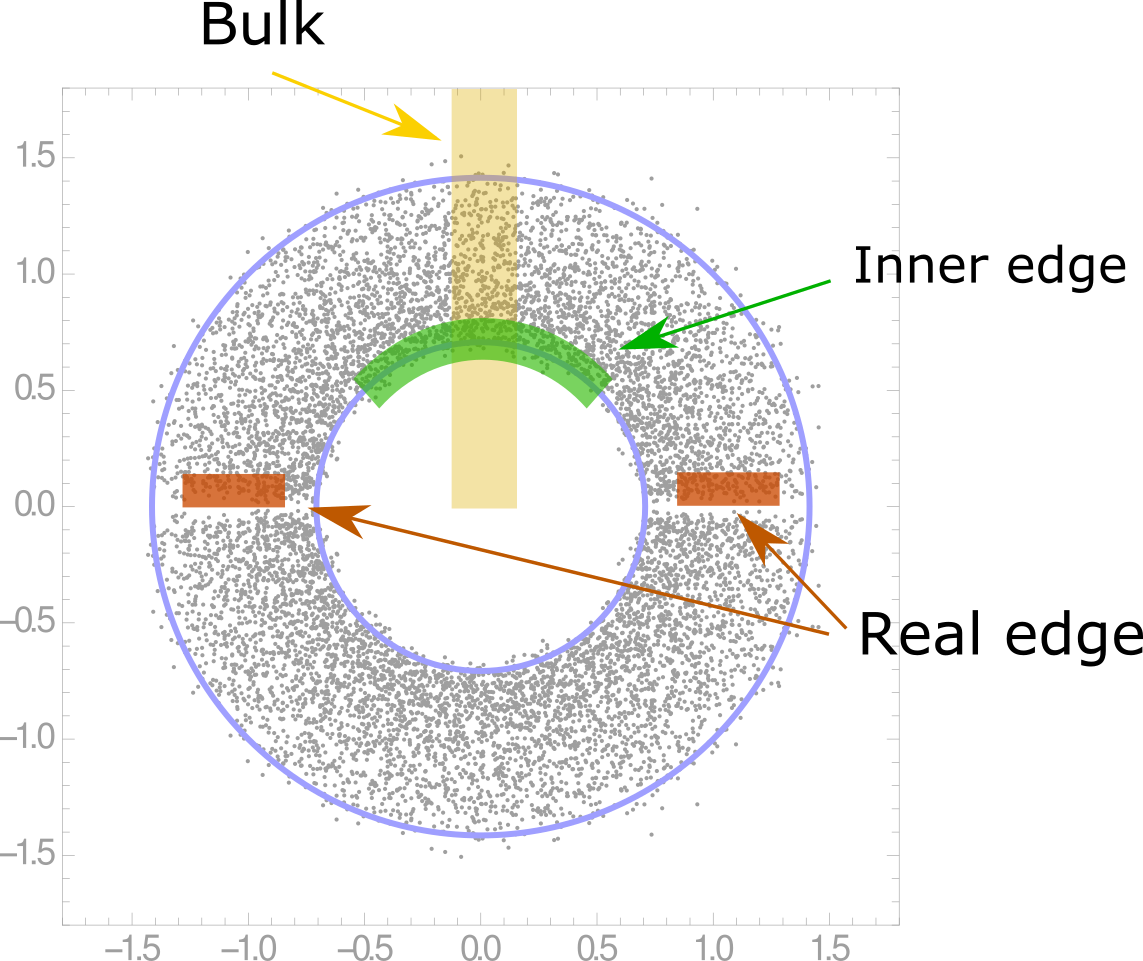}
\end{center}
\caption{\label{f:sampreg}A diagram using  an example eigenvalue plot to show the locations of the samples used to generate the histograms in Figures \ref{f:4bulkedens} (``Bulk''), \ref{f:innerrad} (``Inner edge'') and \ref{f:nearrealevdens} (``Real edge''). This diagram is for illustrative purposes only --- the sampling regions are not drawn to scale.}
\end{figure}

These sampling regions are illustrated in Figure \ref{f:sampreg}.

\end{document}